\documentclass[nohyper,11pt,letterpaper]{JHEP3}

\pdfoutput=1
\usepackage{amssymb,amsmath}
\usepackage{epsfig}
\usepackage{color}

\newcommand\fverb{\setbox\fverbbox=\hbox\bgroup\verb}
\newcommand\fverbdo{\egroup\medskip\noindent%
			\fbox{\unhbox\fverbbox}\ }
\newcommand\fverbit{\egroup\item[\fbox{\unhbox\fverbbox}]}
\newbox\fverbbox

\newcommand{\be}{\begin{equation}}
\newcommand{\ee}{\end{equation}}
\newcommand{\ba}{\begin{eqnarray}}
\newcommand{\ea}{\end{eqnarray}}

\newcommand{\const}{\mbox{const}}
\newcommand{\lap}{\bigtriangleup}
\newcommand{\eq}[1]{Eq.(\ref{#1})}

\newcommand{\BM}[1]{{\mbox{\boldmath $#1$}}}
\newcommand{\pa}{\partial}

\newcommand{\hh}{\, ,\hspace{1cm}}
\newcommand{\hhh}{\, ,\hspace{0.4cm}}

\newcommand{\ins}[1]{{\mbox{\tiny #1}}}

\newcommand{\inds}[1]{{\scriptscriptstyle #1}}

\title{Charged particles in higher dimensional homogeneous gravitational field: Self-energy
and self-force}

\author{Valeri P. Frolov\thanks{E-mail:
vfrolov@ualberta.ca}\, and
Andrei Zelnikov\thanks{E-mail: zelnikov@ualberta.ca}\\
Theoretical Physics Institute, Department of Physics,
University of Alberta, Edmonton, AB, Canada T6G 2E1}

\abstract{A problem of self-energy and self-force for a charged point-like
particle in a higher dimensional homogeneous gravitational field is considered.
We study two cases, when a particle has usual electric charge and a case when it
has a scalar charge, which is a source of a scalar massless minimally coupled
field. We assume that a particle is at rest in the gravitational field, so that
its motion is not geodesic and it has an acceleration $a$ directed from the
horizon. The self-energy of a point charge is divergent and the strength of
the divergence grows with the number of dimensions. In order to obtain a finite
contribution to the self-energy we use a covariant regularization method
which is a modification of the proper time cut-off and other
covariant regularizations. We analyze a relation between the self-energy and
self-force and obtain explicit expressions for the self-forces for the electric
and scalar charge in the spacetimes with the number of dimensions up to eight.
General expressions for the case of higher dimensions are also obtained. We
discuss special logarithmic factors $~\ln a$, which are present both in the
self-energy and self-force in odd dimensions. Finally, we compare the
obtained results with the earlier known results both for the homogeneous
gravitational field and for particles near black holes. }

\preprint{Alberta Thy 13-14}

\begin{document}

\section{Introduction}

There are several `eternal problems' in the theoretical physics,
that have been
discussed already for several decades. The problem of the electromagnetic mass of an
electron is one of them. A charge is a source of the electromagnetic field. The
latter has energy and, when a charged particle moves, it has momentum as well. The
energy of the field contributes to the total energy of the particle, and at least a
part of its proper mass is connected with this contribution. In the classical physics
an electron is a point particle, so that its electromagnetic energy
diverges. In
the simplest case when the charge $e$ is distributed uniformly over the surface of a
sphere of the radius $\varepsilon$ the electromagnetic contribution to the self
energy is
\be\label{EE}
E=e^2/(2\varepsilon)\, .
\ee

The electromagnetic field of a charged particle is distributed in space. Its
configuration, and hence its energy, depends on the boundary conditions. Hence, in a general case the
field contribution to the self-energy besides the local divergent part contains an
additional term which depends on properties of the matter outside
the charge, as well as the boundary conditions. In a general case, the latter
is non-local. If such a contribution depends on the position of the
charge and changes when the charge changes its location, then there exists a non-local
self-force acting on the charge.

Similar effect exists when a charge is located in a spacetime with non-trivial
gravitational field, for example, near a black hole. The black hole metric
acts on the field generated by a charge, in a way,
similar to a dielectric \cite{Landau:1982dva} with the {\it inhomogeneous}
refraction index.
It deforms the electric field so that it is not only
decreasing at
infinity, but takes very special form at the horizon of the black hole, which, for
example, guarantees the regularity of the field in a freely falling frame.
However, using the analogy with a dielectric, it can be quite tricky to arrive to
an intuitive explanation of a correct scaling of the self-force on the
distance from the horizon and even a sign of the effect (see discussion in
\cite{Beach:2014aba}). Concrete calculations are often required to get the
correct answer.
Knowledge of exact solutions in some cases would be of great
luck and help.

When the mass of the black hole is large, the curvature near the event horizon is
small. The horizon becomes practically flat and the gravitational field near it is
approximately static and homogeneous. In other words, the geometry in the vicinity of
the horizon can be approximated by a Rindler metric. If a charge is
located close to the horizon the field near it is similar to the field of a
charge in the Rindler space. However, at far distances from the charge
the difference in  the boundary conditions and the topology of the horizon for
these two cases (the charge near the black hole and in the Rindler spacetime)
becomes important for the self-energy problem. We shall discuss this difference
in the present paper and demonstrate that it is related to the contributions of
zero-modes of the corresponding elliptic field operator.

First calculation of the electromagnetic self-energy of a  charge in a static
homogeneous gravitational field was performed by Fermi in 1921
\cite{Fermi:1921}. In particular, he demonstrated that the `weight' of a system
of electric charges is a product of its electrostatic energy by the
gravitational acceleration, so that the
electromagnetic contribution to the gravitational mass is identical to (properly
calculated) contribution to its inertial mass, as it is required by the equivalence
principle.

For a charge located near a static vacuum black hole the self-energy was calculated
in \cite{Smith:1980tv}.  In the case of an electric charge near a
four-dimensional Schwarzschild black hole the self-force is repulsive.
Later this result was generalized to the case of electric and
scalar charges near Reissner-Nordstr\"{o}m black holes
\cite{Zelnikov:1982in,Zelnikov:1983} and stationary, axisymmetric black holes
\cite{Burko:2001kr}.

Expressions for the finite part of the self-energy in the homogeneous
gravitational field can be obtained by straightforward calculations
\cite{Zelnikov:1983}. The result is very simple
\be\label{exact}
E^{\ins{res}}=-{1\over 2} e^2 a\, .
\ee
The purpose of this paper is to generalize these results to the case when a
charged
particle is in a homogeneous gravitational field in the {\em higher dimensional
spacetime}. There are several reasons why this problem might be interesting.
First of
all, higher dimensional aspects of the high energy physics attracted a lot of
interest in connection with different models of the brane world and large extra
dimensions. In such models the fundamental scale of the quantum gravity can be
in the
energy range of several TeV, which opens an intriguing possibility of the micro
black
hole creation in modern colliders. An interesting question is how the problem of
the
self-energy is modified when the spacetime has one or more extra dimensions.
Another
reason that makes this problem interesting is purely theoretical
curiosity. The electromagnetic field in the higher dimensional spacetime is not
conformal invariant. Moreover, in the case when the spacetime has odd
number of
dimensions, the Huygens-Fresnel principle is violated even in flat
spacetime, namely, the retarded
Green function is not localized on the surface of the past null cones, but has
`tail' inside it. Static Green functions, that are the main topic of this
paper, also reveal specific properties that are alternating with the odd-even
dimensionatity of the spacetime. Recently, Beach et al. \cite{Beach:2014aba}
calculated the self-energy and self-force for an electric and scalar charges near
5-dimensional spherically symmetric vacuum black
hole. They explicitly demonstrates  the presence of a special logarithmic
factor in the expression for the self-force.

An interesting consequence of this phenomena is the following. It can be shown
that the self-energy problem for a point charge in a static $D$-dimensional
spacetime can be
reduced to the calculation of the quantum fluctuations of a scalar field in the
Euclidean $(D-1)$-dimensional theory (see \cite{Frolov:2012zd}). When $D$ is
odd, the
corresponding $(D-1)$-theory possesses quantum anomalies. Their contribution to
the
self-energy of a charge in a Majumdar--Papapetrou class of metrics was analyzed
in
\cite{Frolov:2012xf,Frolov:2012ip}. In three dimensions this anomaly was used
to calculate the self-energy of dipoles \cite{Frolov:2013qia}.

In the present paper we focus on a simpler problem: the self-energy and the
self-force of a point charge in the static homogeneous higher dimensional
gravitational field. We assume that a charge is at rest in such a field. Since
its
worldline is not geodesic, it has acceleration, which we denote by $a$. Let us
emphasize that according to the equivalence principle, this problem is identical
to
a study of the self-force acting on a uniformly accelerated charge in the
Minkowski
spacetime in the absence of the gravitational field.

Because, besides the charge $e$, this problem contains only one dimensional
parameter,
the value of the acceleration, an expected expression for the self-energy and
the
self-force can be easily obtained up to a numerical factor by simple arguments
based
on the dimensional analysis. First of all, let us
notice that the
dimensionality of
the electric charge depends on the number of dimensions. For example, the energy
$E$
of the interaction of two equal charges $e$, and the force $f$ between them in
$D$-dimensional spacetime have
the
form
\be\label{dim}
E\sim {e^2\over R^{D-3}}\, \hh
f\sim {e^2\over R^{D-2}}\, .
\ee
Here $R$ is the distance between the charges.

As we shall see, for a point particle both self-energy and self-force are 
divergent. Based on relations (\ref{dim}) one may arrive to a correct result that the 
leading divergence of these quantities is
\be
E^{\ins{div}}\sim {e^2\over \epsilon^{D-3}}\, \hh
f^{\ins{div}}\sim {e^2\over \epsilon^{D-2}}\, .
\ee
One can define a finite residual part of the self-energy and the self-force, 
which are obtained by subtraction of all the divergences.
Both residual self-energy and self-force
are
also proportional to the charge squared, while $a^{-1}$ has the dimensionality
of the
length\footnote{We use units where $G=c=1$.}. Thus one can write for them
the expressions
\be\label{appr}
E^{\ins{res}}\sim e^2 a^{D-3}\, \hh
{f^{\ins{res}}}\sim e^2 a^{D-2}\, .
\ee
It is easy to see that in the four-dimensional case the expression for
$E^{\ins{res}}$ correctly
reproduces (up to a numerical factor) the exact result \eq{exact}. In the
five-dimensional case, the corresponding exact expressions can be obtained by
taking
proper limit of the results of \cite{Beach:2014aba}. By  the
comparison of the exact results \cite{Beach:2014aba} with \eq{appr} one finds
out that in the five
dimensions these expressions should be modified by a logarithmic term.
Our calculations in the present paper confirm that such a logarithmic
factor appears in all odd-dimensional cases
\be
E^{\ins{res}}\sim e^2 a^{D-3}\ln(a l)\, \hh
f^{\ins{res}}\sim e^2 a^{D-2}\ln(a l)\hh \mbox{$D$ is odd}\,.
\ee

The parameter $l$ which is required in order to make the expression under
the
logarithm dimensionless, is an infrared (IR) cut-off.
It should be emphasized that the case of the homogeneous gravitational field is
not
realistic and is a certain idealization. The gravitational field created by an
extended compact object can be approximated by the homogeneous one only in a
domain,
where
its spatial change (and hence curvature) can be neglected. Similarly, one can
support
 uniform acceleration only for some finite interval of time. One can expect that
the
parameter $l$ reflects the role of these natural infrared cut-offs.

In order to perform calculations of the self-energy for a point charge one
needs
at first regularize its infinities. In higher dimensions this problem becomes
more severe, because divergences are stronger and have a more complicated
structures.
In D dimensions, when $D\ge4$, the leading ultraviolet (UV) divergence of
the self-energy is of the form $e^2/\epsilon^{D-3}$, where $\epsilon$ is the UV
cut-off length parameter. These divergencies can be absorbed into the
renormalization of the local mass of the particle. In four dimensions this mass
renormalization is sufficient to make equations of motion well defined and to 
determine a remaining finite self-energy of the
charged particle, which is the difference of energies of a charged and a
neutral particles of the same renormalized masses. In higher dimensions new
subleading UV divergencies appear which require special treatment.
In a general case these divergences contain the
acceleration and its higher derivatives as well as the curvature and its derivative
(in a case of a curved
spacetime) terms (see, e.g., discussion of the in
\cite{Gal'tsov:2004qz,Galtsov:2007zz}). Simple dimensional analysis shows that
in our case of the homogeneous gravitational field
(constant acceleration)  UV-divergent terms in the self-force have the following
structure
\be\label{PP}
f^{\ins{div}}=e^2 a^{\mu} P\hh
P=\sum_{p=0}^{D-3} c_D^{(p)} {a^{p}\over \epsilon^{D-3-p}}\, .
\ee
In the odd number of spacetime dimensions $D$ the terms in the sum contain also
$\ln(a\epsilon)$ contribution.

One of possible ways to deal with these divergences is to add  corresponding
counter-terms to the action for the particle motion and choose them so that the
divergences can be absorbed by a redefinition of the coefficients of these
counter-terms. These counterterms are necessary to consider even in
flat spacetime and even for a constant acceleration motion
\cite{Kosyakov:1999np}. In six-dimensional curved spacetimes the counterterms
were explicitly calculated in \cite{Galtsov:2007zz}.

For a motion of a particle with the constant acceleration $a$ it is sufficient
to consider only counter-terms of the form
\be\label{count}
S^\ins{counter}=-{1\over 2}\int Q(a) {\bf u}^2 d\tau\hh
Q(a)=\sum_{p=0}^{D-3} C_D^{(p)} a^{p}\, .
\ee
Here ${\mathbf u}$ is the velocity of the particle, $\tau$ is the proper time,
and
$Q(a)$ is a polynomial of the power $D-4$ of the acceleration $a$ (or, in odd
dimensions, a function which is obtained from such a polynomial by including
the $\ln(a\epsilon)$ term).

The variation of this action restricted to the motion with constant $a$ is
\be
{\delta S^\ins{counter}\over \delta x^{\mu}}=a^{\mu}\left[Q(a)-a{\partial
Q(a)\over
\partial a}\right]\, .
\ee
Here the polynomial in the square brackets is of the same order as $Q(a)$. The 
chosen
structure of $S^\ins{counter}$ allows one to include the divergences
\eq{PP} into it by simple redefinition of the coefficients $C_D^{(p)}$ in
$Q(a)$. Note that taking into account logarithmic terms is manageble and does not
complicate the problem.
Let us emphasize that such an approach would inevitably result in a theory
containing higher than second derivatives in the particle equation of motion.
The consistency of such a theory is a complicated problem that we are not able
to discuss here. Let us mention only that if one prefers not to introduce the
counter-terms similar to \eq{count}, one may try to include the divergencies
similar to \eq{PP} into a redefinition of the proper mass of the particle.
However in such a case the self-force of higher-dimensional classical point
charges would become dependent on the composition of extended classical charges.
In other words the self-force can be used as the probe of an internal structure
of extended
sources \cite{Isoyama:2012in}.
In the present paper we adopt the renormalization approach. We shall use a 
modification of the proper time cut-off regularization for calculation of the 
divergencies \eq{PP} and calculate the finite residual part of the self-force 
by subtracting these divergences.

Before describing the structure of the present paper, let us make one more
general
remark. The problem of self-force becomes very popular in connection to the
discussion of motion of compact massive objects (for example small size black holes)
near a large black hole. The methods based on the general theory developed by
DeWitt
and Brehme \cite{DeWitt:1960fc} and their recent modifications are widely used
for this purpose.
We would like to remark that the non-local forces, similar to that we discuss in
this
paper, cannot be found in such calculations.\footnote{Just to give a simple
explanation, let us consider an electric charge in a flat space in the presence
of a
conducting mirror. Since the curvature identically vanishes in the vicinity of
the
charge along all its world-line, all the terms in the Brehme-DeWitt
decomposition are
the same as in the empty space. So that the interaction of the charge with its
`image' is beyond their approximation.}

The paper is organized as follows: In the section~\ref{ElectricSelf-energy} a 
brief
review is given of the general method which we use to compute the self-energy
and self-force for a point electric charge in a static spacetime. In the
section~\ref{ElectricRindler} we apply this
approach and a regularization technique to an electric charge in a
homogeneous gravitational field. In the section~\ref{ElectricResullts} the 
results of
calculations of the self-energy and the self-force of the electric charges in
higher-dimensional spacetimes are collected together. In sections
\ref{ScalarSelf-energy}, \ref{ScalarRindler}, and \ref{ScalarResults} we repeat
the same steps of calculations of the self-energy and the self-force, but in
application to the scalar charges. The Secction~\ref{Discussion} is the summary 
of
the results. In Appendix~\ref{Electric_medium} we provide definitions and a
general derivation of the density of the self-force and self-energy for
electrically and scalar charged media. In Appendix~\ref{GreenFunctions} we
collected the details of calculations of the Green functions expansions, used
in the derivation of our final results for the self-energy and self-force.


\section{An electric charge in a static spacetime}\label{ElectricSelf-energy}

\subsection{Equations}

The action for the Maxwell field in a  $D-$dimensional spacetime has the form
\be
I=-{1\over 16\pi}\int dX\,\sqrt{-g^{\ins{D}}}\,F^{\mu\nu}F_{\mu\nu}+\int
dX\,\sqrt{-g^{\ins{D}}}\,A_{\mu}\,J^{\mu}\, .
\ee
Its variation with respect to the potential $A_{\mu}$ gives the field equation
\be
F^{\mu\nu}{}_{\!;\nu}=4\pi J^{\mu}\, ,
\ee
where $J^{\mu}$ is the current.

Let us consider a  static spacetime and write its metric in the form
\be\begin{split}\label{metric}
&ds^2=g^{\ins{D}}_{\mu\nu}dX^{\mu} dX^{\nu}=-\alpha^2 dt^2+g_{ab}\,dx^a dx^b \hh
 \partial_t\alpha=\partial_t g_{ab}=0\, ,
\end{split}\ee
so that one has
\ba
&&X^{\mu}=(t,x^a)\, ,\ a, b=1,...,D-1\ ,\\
&&g^{\ins{D}}=\det g^{\ins{D}}_{\mu\nu}=-\alpha^2\,g\hhh
g=\det g_{ab}\hhh
g^{\ins{D}}_{00}=-\alpha^2\hhh
g^{\ins{D}}_{ab}=g_{ab}\, .
\ea
This spacetime has the Killing vector $\xi^\mu$ describing the symmetry of the
metric under the time translations. The Killing vector is defined up to an
arbitrary
constant factor, which can be chosen by fixing the norm of the Killing vector
$\xi=\sqrt{-\xi^\mu\xi_\mu}=\alpha(x)$ to be equal to one at some point
$x_\ins{o}$,
that is $\alpha(x_\ins{o})=1$. The proper time of an observer at this point
coincides with the coordinate time $t$.

Later on we consider the energy of a static
point charge located at some point $y$, which does not necessarily coincide with
the
point $x_o$ of the normalization of the Killing vector $\xi^\mu$. At the same
time we
shall use this vector to define the energy of the system. As a result a so
calculated
energy ${\cal E}$ is in fact a function of two variables, ${\mathcal
E}(x_o,y)$.
Similarly one obtains the regularized value and finite residual self-energy
${\mathcal E}_{\epsilon}(x_o,y)$ and ${\mathcal E}^{\ins{res}}(x_o,y)$,
respectively.
These  quantities will be useful for the calculations of the self-force,
because
one is able to find the change of the energy of the charge when its position
$y$
changes, while the normalization point $x_o$ is fixed.
On the other hand, when one discusses the energy of the charged particle, it is
always
possible to choose a position of the normalization point $x_o$ to coincide with
the
position of the charge, $y$, so that one can write
\be
E(y)={\mathcal E}(y,y)\hh E_\epsilon(y)={\mathcal E}_\epsilon(y,y)\hh
E^\ins{res}(y)={\mathcal E}^{\ins{res}}(y,y).
\ee
These are the `energies' evaluated by the observer placed at the position of
the
charge.

For a static source $J^{\mu}=\delta^{\mu}_0 J^0$ the vector potential $A_{\mu}$
can be reduced to the only nontrivial component $A_0$, which obeys the equation
\be\label{Maxwell}
{1\over
\alpha\sqrt{g}}\partial_{a}\left({1\over \alpha}\sqrt{g}\,g^{
ab}\partial_{b}\,A_{0}\right)=4\pi J^0\,.
\ee

Following the paper \cite{Frolov:2012ip} we
introduce another field variable $\psi$ instead of
the electric potential
\be\label{psi}
A_0=-\alpha^{1/2}\,\psi\,.
\ee
Then we can rewrite our problem as that for the scalar field $\psi$ in
\mbox{$(D-1)$-dimensional} space and interacting with the external dilaton
field $\alpha$. The equation for the field $\psi$ is
\be\begin{split}\label{eqpsi}
(\lap  +V)\,\psi&=-4\pi j\hh
\end{split}\ee
Here,
\be
\lap=g^{ab}\nabla_a\nabla_b\,
\ee
is the $(D-1)$-dimensional covariant Laplace operator,
$V$ is the potential, and $j$ is the effective charge density
\be\begin{split}\label{V1}
V&=-{3\over 4}{(\nabla\alpha)^2\over
\alpha^2}+{\lap\alpha\over 2\alpha}
\equiv -\alpha^{1/2}\lap(\alpha^{-1/2})\hh
j\equiv\alpha^{3/2} J^0
 \, .
\end{split}\ee
The field $\psi$ is chosen in such a way that the operator
${\cal O}=(\lap+V)$ is self-adjoint in the space with the metric
$g_{ab}$.

\subsection{Energy}

We define the static Green function ${G}_{00}(x,x')$ as a solution of the
equation
\be\label{electric}
{1\over
\sqrt{g}}\partial_{a}\left({1\over\alpha}\sqrt{g}\,g^{ab}
\partial_{b}{G}_{00}(x,x')\right)={1\over \sqrt{g}}\,\delta(x-x')
\ee
satisfying the properly chosen regularity conditions at the infinity and at the
horizon (if the latter is present). It is easy to check that the static Green
function ${G}_{00}(x,x')$ is the time integral of the $D-$dimensional  retarded
Green function over the whole time $t$ range. Using the static Green function
one
can write the vector potential in the form
\be\label{A0}
A_{0}(x)=4\pi
\int_{\Sigma}dx'\,\alpha(x')
\sqrt{g(x')}\,{G}_{00}(x,
x')\,
J^0(x') \,.
\ee

The energy function ${\mathcal E}$ of a static static charge distribution reads
\be\begin{split}\label{ETmunu}
{\mathcal E}&=-\int_{\Sigma} T_{\mu}^{\nu}\xi^\mu d\Sigma_{\nu}=-\int_{\Sigma}
dx~\sqrt{g}~\alpha
T_{0}^{0}\,.
\end{split}\ee
Substituting here the electromagnetic contribution to the stress-energy tensor
\eq{MaxwellTmunu} one can show that
\be\begin{split}
{\mathcal E}&=\int_{\Sigma} dx\,\sqrt{g}\,\alpha^{-1} g^{ab}\partial_a
A_0\partial_b
A_0\,.
\end{split}\ee
Taking into account the Maxwell equations and boundary conditions at
infinity and at
the horizon we get \cite{Zelnikov:1982in,Zelnikov:1983,Frolov:2012ip}
\be\label{Energy}
{\mathcal E}=-2\pi\int
dx\,dx'\,\sqrt{-g^{\ins{D}}(x)}\,\sqrt{-g^{\ins{D}}(x')}\,J^0(x)\,{G}
_{00}(x,
x')\,J^0(x')\,.
\ee

Using the potential $\psi$ given by \eq{psi} the energy function \eq{Energy} can
be
presented
in the form
\be\begin{split}\label{E2}
{\mathcal E}&={1\over 8\pi}\int
dx\,\sqrt{g}\,g^{ab}\left(\psi_{,a}+{\alpha_{,a}\over
2\alpha}\psi\right)\left(\psi_{,b}+{\alpha_{,b}
\over
2\alpha}\psi\right)
\,.
\end{split}\ee
One can write
\be\begin{split}\label{E3}
{\mathcal E}&=2\pi\int {  dx
dx'}\,\sqrt{g(x)}\sqrt{g(x')}\,j(x)\,{G}(x,x')\,j(x')
\,.
\end{split}\ee
Here ${G}$ is the Green function corresponding to the
operator ${\cal O}=\lap+V$ in \mbox{$(D-1)$-dimensional} space
\be\label{eqG}
(\lap +V)\,{G}(x,x')=-\delta(x,x')\,.
\ee
The Green functions ${G}$ and ${G}_{00}$
are related to each other as
follows
\be
{G}_{00}(x,x')=-
\alpha^{1/2}(x)\,\alpha^{1/2}(x')\,{G}(x,x')\,.
\ee

A point electric charge $e$ moving along the worldline $\gamma$
defined by the equation $X^{\mu}=Y^{\mu}(\tau)$ is described by the
distribution
(see, e.g.,
\cite{Poisson:2011nh,Casals:2012qq,Zimmerman:2014uja})
\be
J^{\alpha}(X)=e\int_{\gamma}d\tau \,
g^{\ins{D}}{}^{\alpha}{}_{\mu}(X,Y(\tau))u^{\mu}\delta^{\ins{D}}(X,
Y(\tau))\,.
\ee
Here $\tau$ is the proper time of the particle, $u^{\mu}=dY^\mu(\tau)/d\tau$ is
the velocity of the particle,
$g^{\ins{D}}{}^{\alpha}{}_{\mu}(X,X')$ is the parallel transport operator, and
$\delta^{\ins{D}}(X,X')$ is the invariant D-dimensional $\delta$-function
\be
\delta^{\ins{D}}(X,X')={1\over \sqrt{-g^{\ins{D}}}}\delta^{\ins{D}}(X-X')\,.
\ee
We are studying static charges in the static spacetimes \eq{metric}. For a
static charge located at a fixed point $y$, the only non-vanishing component of
the
current is
\be\label{J0}
J^0(x)=e{1\over \alpha(y) \sqrt{g(y)}}\,\delta(x-y)\,.
\ee
The rescaled point current \eq{V1} takes the form
\footnote{Charges
are
normalized in such a way that
the interaction energy of two point charges $e_1$ and $e_2$ placed at a
distance $r$ from each other in $D-$dimensional Minkowski spacetime is
\be
{E}={\Gamma\left({D-3\over 2}\right)\over\pi^{D-3\over 2}}
\cdot{e_1 e_2\over r^{D-3}}={4\pi\over (D-3) \Omega_\ins{(D-2)}}\cdot{e_1
e_2\over
r^{D-3}}  \hh
\Omega_n={2\pi^{(n+1)/2}\over \Gamma((n+1)/2)}\,,
\ee
where $\Omega_n$ is the area of $n$-dimensional unit sphere. The interaction
force between charges in $D$ dimensions reads
\be
f={4\pi\over\Omega_\ins{(D-2)}}\cdot{e_1 e_2\over r^{D-3}\,.}
\ee
In the paper \cite{Beach:2014aba} authors work in a different system of
units, such that the force
between two charges $\tilde{e}_1$ and $\tilde{e}_2$ in the $D-$dimensional
Minkowski spacetime is given by
\be
f={\tilde{e}_1 \tilde{e}_2\over r^{D-3}}\,.
\ee
Thus, our normalization of the charges and that of the paper
\cite{Beach:2014aba} are related as
\be
e^2={\Omega_{(D-2)}\over 4\pi}\cdot \tilde{e}^2 \hh
\Omega_2=4\pi\hhh
\Omega_3=2\pi^2\hhh
\Omega_4={8\pi^2\over 3}\hhh
\Omega_5=\pi^3
\,.
\ee
}
\be
j(x)=e {\alpha^{1/2}(y)\over \sqrt{g(y)}}\,\delta(x-y)\,.
\ee
Finally we arrive at the following relation
\be\label{E}
{\mathcal E}=2\pi e^2\,\alpha(y) \,{G}(y,y)\,.
\ee

It was demonstrated in  \cite{Frolov:2012ip} that this relation allows the
following
`elegant' interpretation. Consider \eq{E2} as an action for a quantum field
$\psi$
in a curved $D-1$ dimensional Euclidean space with metric $g_{ab}$ and the
dilaton
field $\alpha$. In this case the Green function $G(x,x')$ in the limit $x'\to x$
is
nothing but the fluctuations of the field $\psi$. Having in mind this
interpretation
in what follows we shall use the notation \be
\langle\psi^2\rangle=\lim_{x,x'\rightarrow y}{G}(x,x')\,.
\ee
Let us notice that one can also write
\be
\alpha(y)\langle\psi^2\rangle=-\lim_{x,x'\rightarrow y}{G}_{00}(x,x')\,.
\ee

Thus one can write \eq{E} in the form
\be\label{renn}
{\mathcal E}=2\pi
e^2\,\alpha \,
\langle\psi^2\rangle\,.
\ee
This representation relates the energy of the point charge
$e$ and quantum fluctuations of the $(D-1)$-dimensional Euclidean quantum field
$\psi$ with the action \eq{E2}. From  $(D-1)$-dimensional point of view $\psi$
is a scalar field.

Of course, the energy of the charge is divergent and it has to be regularized 
(the subscript
$\epsilon$ marks the regularized quantities) either by point splitting
or in any other way
\be\begin{split}
{\mathcal E}_{\inds{\epsilon}}&=2\pi e^2\, \lim_{x,x'\rightarrow y}
\sqrt{\alpha(x)\alpha(x')}\,{G}_{\inds{\epsilon}}(x,x') \\
&=-2\pi e^2\, \lim_{x,x'\rightarrow y}
\,{G}_{\inds{\epsilon}\,00}(x,x')\,.
\end{split}\ee
For our purpose the covariant regularization is
preferable. In this case one can put $x'=x$
while keeping the ultraviolet (UV) regularization parameter $\epsilon$
fixed and only then take a limit $\epsilon\rightarrow 0$.


\subsection{Self-force}

The self-force acting on a static charge in a static spacetime \eq{metric} can
be defined in terms of an
integral of a force density $\mathrm{f}_\alpha(X)$ (in a roman font)
\be\label{forcedensity}
\mathrm{f}_\alpha=-T_{\alpha\beta}{}^{;\beta}
\ee
over the spatial slice $t=\const$ $\Sigma$.
Here $T_{\alpha\beta}$ is the stress-energy tensor of the matter field
including the interaction term with the current, but the contribution of
the mass distribution is excluded.
For a static charge the
force has only spatial components $\mathrm{f}_{\alpha}=(0,\mathrm{f}_a)$. The
integral force acting on the static charge distribution can be defined along
the lines of the articles \cite{Poisson:2011nh,Quinn:1996am,Isoyama:2012in} and
has the form
\be\label{defforce}
f_{a}(y)={1\over\alpha(y)}\int_{\Sigma} g_{a}{}^b(y,x)\,
\mathrm{f}_b(x)\,\alpha(x)\sqrt{g(x)} \,dx\,.
\ee
Here $y$ marks the spatial location of the particle and $g_{a}{}^b(y,x)$ is
the parallel transport operator within the hypersurface $\Sigma$. It's important
to understand that the current density $J^0$, the stress-energy tensor
$T^{\mu\nu}$, the force density $\mathrm{f}_{\alpha}$, etc. are, in fact, the
functions of two points  in $\Sigma$: the point $x$ and the
the position of the particle $y$. For the point charge
$J^0(x)=J^0(x|y)=e\delta(x-y)/(\alpha(x)\sqrt{g(x)})$, and the coordinate $y$
enters
via the argument of the $\delta-$function.

Substituting $T_{\mu\nu}$ for the electric field \eq{MaxwellTmunu} to
\eq{forcedensity} we get
\be
f_{a}(y)={1\over\alpha(y)}\int_{\Sigma} g_{a}{}^b(y,x)\,
F_{a 0}(x)J^{0}(x)\,\alpha(x)\sqrt{g(x)} \,dx\,.
\ee
Then one can use \eq{A0} to obtain
\be
f_{a}(y)={4\pi \over\alpha(y)}\int_{\cal
M}\,dx\,
dx' \sqrt{g(x)}\sqrt{g(x')}\alpha(x)\alpha(x') J^{0}(x) J^{0}(x')
g_{a}{}^b(y,x){\partial\over\partial x^{b}} {G}_{00}(x,x')\,.
\ee
For the point charge  the integration of $\delta-$functions
over the space leads to a simple result
\be\label{fa}
f_{a}(y)=e^2 {4\pi \over\alpha(y)}  {\partial\over\partial x^{a}}
{G}_{00}(x,x')|_{x=y,~x'=y}\,.
\ee
Here the derivative has to be taken before placing points $x$ and $x'$ on the
worldline of the charge. In coincident points
$x=x'=y$ the parallel transport operator reduces to the unit matrix
$\delta_a^b$. One can symmetrize the expression in \eq{fa} to obtain
\be
f_{a}(y)
=2\pi e^2 {1\over\alpha(y)} {\partial\over\partial x^{a}}
\left({G}_{00}(x,x')+{G}_{00}(x',x)\right)|_{x=y,~x'=y}\,.
\ee
We rewrite it as
\be
f_{a}(y)
=2\pi e^2 {1\over\alpha(y)} {\partial\over\partial y^{a}}
{G}_{00}(y,y)\,.
\ee

This expression is formal because it is divergent and has to be properly
regularized.
Regularization of the self-force
can be performed by the same methods as those of the self-energy.
Then the regularized self-force acting on the charge
can be written in terms of a variation of the self-energy function over the
position
of the charge.
\be\begin{split}\label{fepsilon}
f_{\inds{\epsilon}}{}_{a}
=2\pi e^2 {1\over \alpha(y)}
\partial_{a}\,{G}_{\inds{\epsilon}00}(y,y)
=-2\pi e^2 {1\over \alpha(y)}
{\partial\over\partial y^{a}}\left(\alpha(y){G}_{\inds{\epsilon}}(y,
y)\right)=-{1\over
\alpha}{\partial\over\partial y^{a}}{\mathcal E}_{\inds{\epsilon}}\,.
\end{split}\ee
Note that $\alpha=1$ at the position of the observer $x_\ins{o}$ rather than the
charge. Therefore. if we want to evaluate the self-force at the position $y$ of
the
charge, then we can put $\alpha(y)=1$, but only at the very end of computations
after taking the derivative over $y$.


\section{Electric charge in the Rindler spacetime}\label{ElectricRindler}

\subsection{Static Green function}

The $D-$dimensional Rindler metric which describes a static homogeneous
gravitational
field reads
\be\begin{split}\label{Rindler}
ds^2&=-a^2 z^2 dt^2+dz^2+d\boldsymbol{x}_{\perp}^2\hhh
d\boldsymbol{x}_{\perp}^2=\delta_{ij}dx^i dx^j\hh i,j=2,\dots, (D-1)\,,\\
\alpha&=az\hh
x\equiv x^a=(z,x^i)\,.
\end{split}\ee
We consider a particle at rest at the proper distance $z$ from the horizon.
Using the
translation invariance of the metric one can always choose $x^i=0$. In what
follows
we assume this choice.

The particle at rest at $z=\const$ has a velocity
\be
u^{\alpha}={1\over az}\,\delta^{\alpha}_{t}\,,
\ee
and it experiences a constant acceleration
\be
w^{\alpha}={1\over z}\,\delta^{\alpha}_z\,.
\ee
We remind that $t$ is a proper time of an observer at rest at $z=a^{-1}$ and, 
hence, at this point
\be
w^{\alpha}=a\,\delta^{\alpha}_z\,.
\ee

Our next goal is to find the static electromagnetic Green function in the 
Rindler spacetime. In
a general case one can add  to it a solution of the corresponding homogeneous
equation. This ambiguity is fixed by a proper choice of the boundary
conditions. In our case the homogeneous equation is
\be
(\lap+V)\psi =\left[\partial_z^2+\partial_{\boldsymbol{x}_{\perp}}^2-{3\over
4z^2}\right]\psi=0\, .
\ee
A solution can be decompose into modes
\be
\psi\sim \exp(i\boldsymbol{k}_{\perp}\boldsymbol{x}_{\perp})Z(z)\, ,
\ee
where $Z$ is a solution of the equation
\be
{d^2 Z\over dz^2}-|\boldsymbol{k}_{\perp}|^2 Z -{3\over 4z^2}Z=0\, .
\ee
This equation has two singular points: horizon $z=0$ and infinity $z=\infty$.
Its solution has two arbitrary constants. For $|\boldsymbol{k}_{\perp}|\ne 0$
one of them is `killed'  by the requirement that the solution is finite at
the infinity. Near the horizon one has
\be
Z\sim C_0 z^{-1/2}+C_1 z^{3/2} +\, .
\ee
Thus we obtain
\be
A_0=-\alpha^{1/2}\psi\sim -a^{1/2} (C_0 + C_1
z^{2})\exp(i\boldsymbol{k}_{\perp}\boldsymbol{x}_{\perp})\, .
\ee
Keeping the leading term at the horizon one has for $|\boldsymbol{k}_{\perp}|\ne
0$
\be\label{FF}
F_{\mu\nu}F^{\mu\nu}\sim z^{-2} A_{0,\boldsymbol{x}_{\perp}}^2\sim C_0^2
\boldsymbol{k}_{\perp}^2 z^{-2}\, .
\ee
Thus a regularity of the electromagnetic field at the horizon implies that
$C_0=0$. To summarize, in a general case a homogeneous solution dependent on
$\boldsymbol{x}_{\perp}$ cannot be both regular at the horizon and restricted
at infinity. For $|\boldsymbol{k}_{\perp}|= 0$ the regularity of the field
strength at the horizon does not restrict constants $C_0$ and $C_1$. 
Such a solution is, in fact, a zero mode of our field operator. The
remaining freedom is a choice of the constant potential value $A_0$ at the
spatial infinity. But one can always put it equal to zero without changing the
strength of the field. Such a solution is unique. This simple analysis shows
that if only one finds the static Green function for our problem, which is
regular at the horizon and vanishing at the infinity, this solution is unique.

For the Maxwell field we have to find the regularized Green for the operator 
\eq{eqpsi} with the potential \eq{V1}. In the
Rindler spacetime $V=-{3/(4 z^2)}$ and
\be
(\lap+V)\,G(x,x')
=\left[\partial_z^2+\partial_{\boldsymbol{x}_{\perp}}^2-{3\over
4z^2}\right]G(x,x')=-\delta(x,x')\, .
\ee
We construct the Green function by using the heat kernel 
$K(s|x,x')$, which is the solution the equation
\be\label{eqK}
\left[-{\partial\over\partial s}+\lap+V\right]K(s|x,x')=0
\ee
with the boundary conditions
\be\label{eqK0}
K(0|x,x')=\delta(x,x')\,.
\ee
For the Rindler metric \eq{Rindler} $\lap$ is a flat $(D-1)$-dimensional
Laplace
operator. The static Green function
\be
{G}(x,x')=\int_0^{\infty}ds\,{K}(x,x')\,.
\ee

The static Green function in coincidence limit and, hence, the self-energy
function diverges. The standard way of extracting the UV
divergences is to regularize the heat kernel. In general the regularized heat
kernel  can be obtained by multiplying the heat kernel by
a weight function $\rho(s,\epsilon)$, which vanishes for
$s\ll\epsilon^2$ and is equal to one for $s\gg\epsilon^2$. For example, a
proper time cut-off regularization corresponds to a choice
$\rho=\theta(s-\epsilon^2)$. For our calculations it is more convenient to
choose
\be
\rho(s,\epsilon)=\exp(-{\epsilon^2/4s})\,,
\ee
\be\label{heatkernel}
K_{\inds{\epsilon}}(s|x,x')=\rho(s,\epsilon)\,K(s|x,x')\,.
\ee
The corresponding regularized static Green function reads
\be\label{GKds}
G_{\inds{\epsilon}}(x,x')=\int_0^{\infty}ds\,K_{\inds{\epsilon}}
(s|x,x')\,.
\ee
Note that this regularization is covariant and can be performed originally in
$D$ dimensions for arbitrary spacetime. The result of an integration of this
heat kernel over the time $t$ is consistent with the
static heat kernel $K_{\inds{\epsilon}}(s|x,x')$.

One can check that the solution of \eq{eqK}-\eq{eqK0} for the regularized static
heat kernel is
\be\label{K0}
K_{\inds{\epsilon}}=\sqrt{zz'}\,{2\pi\over(4\pi s)^{D/2}}\,
e^{\displaystyle{-{z^2+z'{}^2+\boldsymbol{x}_{\perp}^2+\epsilon^2\over
4s}}}\,I_1\left({}zz'\over
2s\right)\,,
\ee
where $I_1$ is the Bessel function and
\be
\boldsymbol{x}_{\perp}^2 \equiv \delta_{ij}(x^i-x'{}^i)(x^j-x'{}^j)\,.
\ee
The choice of the Bessel function $I_1$ is dictated by the condition that in
the limit of small $s$ the heat kernel \eq{K0} tends to the regularized heat
kernel in the $(D-1)$-dimensional flat space
\be\label{K0small}
K_{\inds{\epsilon}}\Big|_{s\rightarrow 0}\rightarrow {1\over(4\pi s)^{{D\over
2}-1}}\,
e^{\displaystyle{-{(z-z')^2+\boldsymbol{x}_{\perp}^2+\epsilon^2\over
4s}}}\,.
\ee

The regularized Green function is the integral \eq{GKds} of the regularized
heat kernel over the proper time
\be\label{MaxwellG}
{G}_{\inds{\epsilon}}={\sqrt{zz'}\over
2\pi^\beta}\,I_{1}^{\beta}\hh
\beta={{D\over 2}-1}\,,
\ee
and
\be
{G}_{\inds{\epsilon}}{}_{00}=-a{zz'\over 2\pi^\beta}
\,I_{1}^{ \beta}\,,
\ee
where we defined
\be\begin{split}
I^\eta_\nu&=\int_0^{\infty}u^{\eta-1}\,e^{-pu}I_\nu(cu)\,du\,.
\end{split}\ee
Here are a few useful forms of the integral expressed in terms of the
hypergeometric function $F$ or the associated Legendre function
$P_{\sigma}^{\rho}$
\be\begin{split}\label{Prudnikov}
I^\eta_\nu&=p^{-\eta-\nu}\left({c\over
2}\right)^{\nu}{\Gamma(\eta+\nu)\over\Gamma(\nu+1)}\,{F}\left({
\eta+\nu\over 2},{\eta+\nu+1\over 2};\nu+1;{c^2\over p^2}\right)
\\
&=(p-c)^{-\eta-\nu}\left({c\over
2}\right)^{\nu}{\Gamma(\eta+\nu)\over\Gamma(\nu+1)}\,{F}\left(
\eta+\nu,\nu+{1\over 2};2\nu+1;-{2c\over p-c}\right)
\\
&=e^{-\pi\nu i/2}\Gamma(\eta+\nu)(p^2-c^2)^{-\eta/2}\,P^{-\nu}_{\eta-1}
\left( { p\over\sqrt { p^2-c^2 } } \right)\,,
\end{split}\ee
where
\be
u={1\over 4s}\hh p=z^2+z'{}^2+\boldsymbol{x}_{\perp}^2+\epsilon^2\hh c=2zz'\,.
\ee
Let $R$ and $\bar{R}$ be the distances from the observation point to the charge and
its image correspondingly
\be\label{R}
R^2=(z-z')^2+\boldsymbol{x}_{\perp}^2 \hh
\bar{R}^2=(z+z')^2+\boldsymbol{x}_{\perp}^2\,.
\ee
The combination $p-c=R^2+\epsilon^2$ entering the Green function controls its
singular behavior in the limit of coincident points. Using the relations
\be
p={1\over 2}(\bar{R}^2+R^2)+\epsilon^2   \hh
c={1\over 2}(\bar{R}^2-R^2)
\ee
one can express the heat kernel and the Green function in terms of $R$ and
$\bar{R}$.

Thus we can write the regularized static Green function in the form
\be\label{G00}
{G}_{\inds{\epsilon}}{}_{00}=-a{\Gamma\left(\beta+1\right)
\over 2\pi^{\beta}}
{(zz')^2\over (R^2+\epsilon^2)^{\beta+1}}\,{F}
\left(\beta+1,{3\over 2};3;-{4zz'\over R^2+\epsilon^2}\right)
\,,
\ee
One can check that both
${G}_{\inds{\epsilon}}{}_{00}(x,x')$ and
${G}_{\inds{\epsilon}}(x,x')$ vanish when one of the
arguments, e.g., $x$, is placed on the horizon. It means that they satisfy
zero Dirichlet boundary conditions on the Rindler horizon.

To obtain the general solution for the Green function $G_{00}$ one should remove
regularization (put $\epsilon=0$) in
\eq{G00} and add a symmetric in $z$ and $z'$ zero-mode solutions
$
C_0+C_1\,(z^2+z'{}^2)+C_2\,z^2z'{}^2
$
of a homogeneus equation of the Maxwell equation.

This general solution
\be\label{A00}
G_{00}=-a{\Gamma\left({{D\over 2}}\right)
\over 2\pi^{{D\over 2}-1}}
{(zz')^2\over R^{D}}\,{F}
\left({{D\over 2}},{3\over 2};3;-{4zz'\over
R^2}\right)+C_0+C_1\,(z^2+z'{}^2)+C_2\,z^2z'{}^2
\ee
is parametrized by three arbitrary constants $C_0, C_1, C_2$, which are to
be
fixed by the boundary conditions at the horizon and at infinity. In our
particular case we require $G_{00}$ to be finite, when either $z$ or $z'$ are
on the horizon,
and $G_{00}$ to vanish at infinity. It leads to the choice $C_0$, $C_1=C_2=0$.

The static Green functions for particular dimensions of the spacetime are given
in the Appendix~\ref{GreenFunctions}. Their asymptotics at
$\epsilon\rightarrow 0$ can be found in Appendix~\ref{MaxwellField}.

\subsection{Near-horizon limit of the Schwarzschild black hole}

The geometry of a static black hole near the horizon can be approximated by the
Rindler metric. If a charge is close to the horizon its field in the Rindler
domain can be described by the above constructed solution. This means that this
solution can be obtained as a special limit of the field created by the charge
in the black hole geometry. We illustrate this by a simple example of a four
dimensional Schwarzschild metric, where the exact solution for the field of a
point charge is known.
The static Green function near a Schwarzschild black hole of mass $M$ is
is  \cite{Copson01031928,Linet:1976sq}
\be\nonumber
G^\ins{Schw}_{tt}=-{M\over 4\pi r r'}-{(r-M)(r'-M)-M^2\cos\theta\over 4\pi rr'
\sqrt{(r-M)^2+(r'-M)^2}-2(r-M)(r'-M)\cos\theta -M^2\sin^2\theta}\,.
\ee
Here the first term (the Linet term \cite{Linet:1976sq}) is singular
only inside the black hole. It describes the potential of a weakly charged
black hole. This term $-M/(4\pi rr')$ is important to keep the total charge
of the black hole equal to zero and
corresponds to the zero mode which is finite on the horizon and vanishes at
infinity.

In the near horizon limit this term boils down to a trivial constant zero mode
in the Rindler space. Near the horizon the Schwarzschild radial and angle
coordinates are related with the proper distance $z$ to the horizon and
transverse coordinate $x_\perp$ as
\be
r=2M\left(1+{z^2\over 16 M^2}-{z^4\over 768 M^4}\right)+O(M^{-6})\,.
\ee
\be
\cos\theta=1-{\boldsymbol{x}^2_\perp\over 8M^2}+O(M^{-4})
\ee

The Rindler limit can be obtained by taking the limit of large 
$M$ provided one considers the domain near the horizon, where 
the proper distances from the horizon to 
the charge and the observation point are kept fixed and finite. In this limit 
the coordinate time 
$x^0$ is defined as the red-shifted Schwarzschild time $t$ rescaled in such a 
way that 
the timelike Killing vector is normalized to unity when the proper distance to 
the horizon is equal to $a^{-1}$.

Then one gets
\be
G^\ins{Schw}_{00}\approx-{a\over 8\pi}\,{(R+\bar{R})^2\over
R\bar{R}}\,,
\ee
where $R$ and $\bar{R}$ are given by \eq{R}. 

Our solution \eq{A00} in four-dimensional Rindler space takes the form
\be
G_{00}=-{a\over 8\pi}\,{(R-\bar{R})^2\over R\bar{R}}\,.
\ee
The difference between them is, evidently, the constant zero mode contribution
\be
G^\ins{Schw}_{00}=G_{00}-{a\over 2\pi}\,.
\ee
Similar calculations can be done for the scalar Green functions. In that case
no extra zero mode contributions appear. The difference of the Green
functions for
the black hole and Rindler spacetimes reflects the difference of their geometry
and topology of the horizon.


\section{Electromagnetic self-energy and self-force:
Results}\label{ElectricResullts}

\subsection{Divergent part of the static Green function}

In order to derive renormalized self-energy function
one has to subtract the UV divergent parts from the Green function
\eq{MaxwellG}. Their local structure is given by the Hadamard expansion. They
can also be deduced from the heat kernel expansion in powers of the proper time
$s$. The generic structure of these divergences is
\be
K_{\inds{\epsilon}}^{\ins{div}}(s|x,x')={e^{-{{
2\sigma(x,
x')+\epsilon^2\over 4s}-\lambda^2 s}}\over (4\pi s)^{(D-1)/2}}
[a_0(x,x')+s a_1(x,x')+\dots+s^{[(D-3)/2]} a_{[(D-3)/2]}(x,x')]\,.
\ee
Here $a_n(x,x')$ are the Schwinger-DeWitt coefficients.
Because we are dealing with the regularized version of the heat kernel, we can
safely take first a coincidence limit $x=x'$ and then integrate the obtained
expression over the proper time $s$
\be
{G}_{\inds{\epsilon}}^{\ins{div}}(x,x)=\int_0^{\infty}
ds\,K_{\inds{\epsilon}}^{\ins{div}}(s|x,x)\,,
\ee
and finally take an asymptotic of the result at small values of the UV
cut-off $\epsilon$. When $x=x'$
\be\label{DeWitt}
K_{\inds{\epsilon}}^{\ins{div}}(s|x,x)={e^{-{{
\epsilon^2\over
4s}-\lambda^2
s}}\over (4\pi s)^{(D-1)/2}}
[a_0(x,x)+s a_1(x,x)+\dots+s^{[(D-3)/2]} a_{[(D-3)/2]}(x,x)]\,.
\ee
Here $\lambda$ is an arbitrary IR cut-off parameter.
In the case of the electric charge in the Rindler spacetime the
first three Schwinger-DeWitt  coefficients are
\be
a_0(x,x)=1\hhh
a_1(x,x)=V=-{3\over 4\,z^2}\hhh
a_2(x,x)={1\over 2}V^2+{1\over 6}\lap V=-{15\over 32\,z^4}\,.
\ee

Now expanding ${G}_{\inds{\epsilon}}(x,x)$ at small $\epsilon$ and
subtracting the ${G}_{\inds{\epsilon}}^{\ins{div}}(x,x)$ we get
\be
\langle\psi^2\rangle={G}_{\inds{\epsilon}}(y,y)-{G}_{
\inds{\epsilon}}^{\ins{div}}(y,y)\, .
\ee

\subsection{Self-force}\label{MaxwellSelf-force}

Substitution of the results of the above  calculations for different spacetime
dimensions (see Appendix~\ref{MaxwellField}) into \eq{E}, \eq{fepsilon} allows 
one to find both the divergent and the finite residual parts of the
self-energy and the self-force. The results can be summarized as 
follows.
One starts with the equation of motion of a neutral particle with mass $m_0$
\be\label{eom}
m_0 a_{\mu}= F_{\mu}^\ins{ext}\, ,
\ee
where $F_{\mu}^\ins{ext}$ is an external force. When the particle has an 
electric charge $e$ this equation is modified and takes the form
\be\label{modeq}
m_0 a_{\mu}= F_{\mu}^\ins{ext}+f_{\mu} \, ,
\ee
where $f_{\mu}$ is an additional `self-force' proportional to $e^2$. The 
calculations show that it can be presented in the form
\be\label{ff}
f_{\mu}=f_{\mu}^\ins{div}+f_{\mu}^\ins{res}\, .
\ee
Here $f_{\mu}^\ins{div}$ is a divergent in the limit $\epsilon\to 0$ part of the 
self-force. It has
the following structure
\be\label{PP}
f_{\mu}^\ins{div}=e^2 a_{\mu} P(a)\hh
P(a)=\sum_{k=0}^{k_D} c_D^{(2k)} {a^{2k}\over \epsilon^{D-3-2k}}\, .
\ee
Here $D$ is the number of spacetime dimensions and $k_D=(D-3)/2$ for odd $D$ and 
$k_D=D/2-2$ for even $D$.
For odd $D$ in the last term of this expression one should substitute
\be
{1\over \epsilon^0}\to \ln(\epsilon\tilde{\lambda})\hh \tilde{\lambda}=\lambda 
e^{\gamma}/2\, ,
\ee
where $\lambda$ is the infrared cut-off. The numerical coefficients $c_D^{(2k)}$ 
are 
given below.

\bigskip

\hskip 2cm
\begin{tabular}{|r|l|l|l|}
  \hline
  $D=4 \phantom{\displaystyle{1\over1}}$& $c_4^{(0)}=-{1\over 2}$& & \\
  \hline 
  $D=5\phantom{\displaystyle{1\over1}}$ & $c_5^{(0)}=-{1\over 2\,\pi}$& 
$c_5^{(2)}={3\over 16\,\pi}$&\\
  \hline 
  $D=6\phantom{\displaystyle{1\over1}}$ & $c_6^{(0)}=-{1\over 4\,\pi}$& 
$c_6^{(2)}=-{3\over 32\,\pi}$&\\
  \hline
  $D=7\phantom{\displaystyle{1\over1}}$ & $c_7^{(0)}=-{1\over 2\,\pi^2}$& 
$c_7^{(2)}=-{3\over 32\,\pi^2}$&$c_7^{(4)}={45\over 512\,\pi^2}$\\
  \hline
  $D=8\phantom{\displaystyle{1\over1}}$ & $c_8^{(0)}=-{3\over 8\,\pi^2}$& 
$c_8^{(2)}=-{3\over 64\,\pi^2}$&$c_8^{(4)}=-{45\over 
1024\,\pi^2}$\\
  \hline
\end{tabular}
\bigskip

It should be emphasized that $f_{\mu}^\ins{div}$ is determined by local terms 
in the 
Hadamard expansion of the Green function. We denote by $f_{\mu}^\ins{res}$ the 
finite 
part of the self-force which is the residue obtained by subtracting the local 
divergent terms from $f_{\mu}$. This part of the force depends on the boundary 
conditions and, hence, in a general case is non-local. Our calculations give 
\be\label{fmures}
f_{\mu}^\ins{res}=e^2 a_{\mu} a^{D-3}[A_D +B_D 
\ln(8\tilde{\lambda}/a)]\, .
\ee
For even $D$ the logarithmic terms are absent and one has $B_D=0$.
The coefficients $A_D$ and $B_D$ (for odd dimensions) are summarized 
below.

\bigskip

\hskip 2cm
\begin{tabular}{|r|l|l|l|}
  \hline
  $D=4 \phantom{\displaystyle{1\over1}}$& $A_4=0$& ${B}_4=0$ \\
  \hline 
  $D=5\phantom{\displaystyle{1\over1}}$ & $A_5={11\over 32\,\pi}$& 
${B}_5=-{3\over 16\,\pi}$\\
  \hline 
$D=6\phantom{\displaystyle{1\over1}}$ & $A_6=0$& 
${B}_6=0$\\
  \hline 
$D=7\phantom{\displaystyle{1\over1}}$ & $A_7={381\over 2048\,\pi^2}$& 
${B}_7=-{45\over 512\,\pi^2}$\\
  \hline 
$D=8 \phantom{\displaystyle{1\over1}}$& $A_8=0$& ${B}_8=0$ \\
  \hline 
\end{tabular}
\bigskip

The leading divergence of $P(a)$ is $\sim  {c_D^{(0)}/\epsilon^{D-3}}$ 
and it does 
not depend on the acceleration. One can absorb it into $m_0$, by redefining the 
mass 
\be\label{m0m}
m_0\to m=m_0-e^2 {c_D^{(0)}\over \epsilon^{D-3}}\, .
\ee
In the four dimensional spacetime this is the only divergent term in the 
expression for $f_{\mu}^\ins{div}$, so that one can say that the problem of 
divergence of 
the self-force can be solved by a standard method of the mass renormalization, that 
is by redefining the universal coupling constant in the action for the particle 
motion. After this one usually say that we must use the action where $m_0$ is 
substituted by $m$, and the value of this parameter $m$ should be determined from 
observations.

In the higher dimensions one certainly meets a new problem: there exist subleading 
divergences in the self-force and they depend on the choice of the solution (on 
the parameter of the 
acceleration). One may say that these divergences can also be absorbed into a 
redefinition of the mass of the particle, so that after such a redefinition the 
equation (\ref{eom}) would take the form 
\be
m(a) a_{\mu}=F_{\mu}^\ins{ext}+f_{\mu}^\ins{res}\, .
\ee
However, in order to obtain such an equation from the least action principle 
one should modify the action for a free 
particle. Namely instead of the standard action for a neutral particle \be
S_0=-{m_0\over 2}\int d\tau \BM{u}^2\, ,
\ee
one should consider the action
\be\label{genact}
S=S_0-{1\over 2}\int d\tau Q(a) \BM{u}^2\hh Q(a)=\sum_{k=0}^{k_D} C_D^{(2k)} 
a^{2k}\, .
\ee
Here $\BM{u}$ is the velocity of the particle.
For odd $D$ the term $C_D^{(D-3)} a^{D-3}$ with highest power of 
$a$ 
should be 
replaced by $[C_D^{(D-3)} 
+\tilde{C}_D^{(D-3)}\ln(8\tilde{\lambda}/a)]a^{D-3}$. 
The coefficient $C_D^{(0)}$ is nothing but the change of the mass of the 
particle. The variation of this action results in the following new form of the 
equation of motion
\be\label{QQ}
m_0 a_{\mu}+\left(Q-a{\pa Q\over \pa 
a}\right)a_{\mu}+\ldots=F_{\mu}^\ins{ext}+f_{\mu}^\ins{res}\, .
\ee
The dots denote terms which contain $\dot{a}^{\nu}$ and $\ddot{a}^{\nu}$. These 
terms vanish for the motion with constant acceleration and the equation takes 
the 
form \be\label{QQQ}
m_0 a_{\mu}=\left(a{\pa Q\over \pa 
a}-Q\right)a_{\mu}+F_{\mu}^\ins{ext}+f_{\mu}^\ins{res}\,.
\ee
By comparing this equation with \eq{ff} and \eq{PP} one finds that they are the 
same 
if
\be
a\,{\pa Q\over \pa a}-Q=e^2 P(a)\, ,
\ee
or, what is equivalent, if one chooses
\be\label{CC}
C_D^{(2k)}={e^2\over 2k-1} {c_D^{(2k)}\over \epsilon^{D-3-2k}}\, .
\ee

Let us make a few comments connected with possible 
interpretations of the obtained results. One can try to interpret \eq{QQ} as 
follows. After absorbing the term $Q(a=0)$ into a redefinition 
(renormalization) 
of the mass, one may insist that the rest divergent terms in the right-hand 
side 
of \eq{QQQ} describe a {\em real physical force} acting on an accelerated 
charged particle in the higher dimensions. However, there is another option, 
which from our point of view is more preferable. Namely, one can assume that in 
order to have a consistent theory of a particle in the higher dimensions one 
needs to start with a generalized action of the form  \eq{genact} with 
arbitrary 
coefficients $C_D^{(2k)}$. In such a case, in the presence of charge the 
subleading divergent terms in \eq{modeq} can be absorbed by redefinition of 
these coefficients by adding the terms of the form \eq{CC}. It should be 
emphasized that this renormalization procedure does not require the knowledge 
of 
the state of motion (acceleration $a$). After such a redefinition is performed, 
following the standard  renormalization procedure adopted in the quantum field 
theory, one may consider $C_D^{(2k)}$ as the finite renormalized coupling 
constants, the value of which should be found from experiments. This approach was 
advocated, for example, by Galtsov in \cite{Gal'tsov:2004qz,Galtsov:2007zz}. In such 
a case the finite 
non-local contribution to the force $f_{\mu}^\ins{res}$ can be interpreted as 
a finite self-force.

\subsection{Self-energy}

After this detailed dicussion of the self-force let us make a few brief 
comments on the residual (renormalized) value of the self-energy.
In four dimensions we reproduce the old result for the self-energy function
\cite{Zelnikov:1983}
\be
{\mathcal E}^{\ins{res}}=-{e^2\,a\over 2}\hh D=4\,.
\ee
We present here also the results for $E^\ins{res}$. The residual terms of the 
self-energy function in other dimensions can be obtained by using Appendix 
\ref{MaxwellField}.

In even-dimensional spacetimes higher than four dimensions the energy function
seems
to vanish, though concrete computations we performed for $D=6$ and $D=8$,
\be
{\mathcal E}^{\ins{res}}=0\hh D=6,8\,.
\ee
In odd-dimensional spacetimes  $D=5,7$ we obtain
\be
{\mathcal E}^{\ins{res}}=-{3\,e^2\,a\over 16\pi z}
\left[\ln\left({4\lambda z}\right)+\gamma-{5\over 6}\right]\hh D=5\,,
\ee
\be
{\mathcal E}^{\ins{res}}=-{15\,e^2\,a\over 512\pi^2 z^3} \left[\ln\left({
4\lambda z}\right)+\gamma-{107\over 60}\right] \hh D=7\,.
\ee
Let us remind that the self-energy functions ${\mathcal E}^{\ins{res}}(x_o,y)$
depend on two points, a position of the charge $y^a$ and the point of
normalization
of the Killing vector $x_o^a$. In the above expressions we use the following
choice:
$x_o^a=(a^{-1},x^i=0)$ and $y^a=(z,y^i=0)$. Hence the energy functions are
functions of
two parameters $a$ and $z$. For the self-energy one has
${E}^{\ins{res}}(a^{-1})={\mathcal
E}^{\ins{res}}(a^{-1},a^{-1})$. The expression for the self-force of the
charge,
given by
\eq{fepsilon} calculated at the point $z=a^{-1}$.



\subsection{Summary of results}

The results of the calculations of the residual self-energy ${E}^{\ins{res}}$ 
and self-force ${f}_a^{\ins{res}}$ can be
summarized as follows:
\begin{itemize}
\item Four dimensions. Self-energy is negative and constant. The self-force
vanishes.
\be
{E}^{\ins{res}}=-{e^2\,a\over 2} \hh {f_z^{\ins{res}}}=0\,.
\ee
\item Five dimensions.
\be
{E}^{\ins{res}}=-{3\,e^2\,a^2\over 16\pi}
\left[\ln\left({4\lambda \over a}\right)+\gamma-{5\over 6}\right]\,,
\ee
\be
{f_z^{\ins{res}}}=-{3\,e^2 a^3\over 16\pi}
\left[\ln\left({4\lambda\over a}\right)+\gamma-{11\over 6}\right]\,.
\ee
\item Six dimensions
\be
{E}^{\ins{res}}=0 \hh
{f_z^{\ins{res}}}=0\,.
\ee
\item Seven dimensions.
\be
{E}^{\ins{res}}=-{15\,e^2\,a^4\over 512\pi^2} \left[\ln\left({
4\lambda \over a}\right)+\gamma-{107\over 60}\right]\,,
\ee
\be
{f_z^{\ins{res}}}=-{45\,e^2 a^5\over 512\pi^2} \left[\ln\left({
4\lambda \over a}\right)+\gamma-{127\over 60}\right]\,.
\ee
\item Eight dimensions
\be
{E}^{\ins{res}}=0\hh
{f_z^{\ins{res}}}=0\,.
\ee
\end{itemize}
The invariant self-force is equal to the absolute value of the $z$-component of
the
self-force ${f^{\ins{res}}}=\sqrt{f_a^\ins{res}
f^{\ins{res}\,a}}=|f_z^{\ins{res}}|$.

One can see that residual the self-force in even dimensions vanishes. We 
obtained this result
for $D\le 8$, however one can make a conjecture that this result is valid 
in any even dimensional spacetimes with a static homogeneous gravitational 
field. Let us also notice that
in odd dimensions the residual self-energy and the residual self-force depend on
the IR cutoff $\lambda=1/l$ which has the dimensionality of inverse length. A
similar logarithmic terms are also present in the expression for the self-force
of a charge near a five-dimensional static black hole (see
discussion in \cite{Beach:2014aba}).


\section{A scalar charge in a static spacetime}\label{ScalarSelf-energy}

\subsection{Equations}

Let us consider now a self-energy of a scalar charge in a static
spacetimes.
A minimally coupled massless scalar field $\varPhi$
is described by an action
\be
I=-{1\over 8\pi}\int
dX\,\sqrt{-g^{\ins{D}}}\,\varPhi^{;\mu}\varPhi_{;\mu}+\int
dX\,\sqrt{-g^{\ins{D}}}\, J \varPhi\,.
\ee
It obeys the equation
\be\label{BoxPhi}
\Box\,\varPhi=-4\pi J\, .
\ee
The energy function ${\mathcal E}$ of a static configuration of the scalar
field
is
given by the
integral \eq{ETmunu} of the stress-energy tensor of the scalar field
\eq{ScalarTmumu}. Taking into account only the scalar field contribution, we
have
\be
T_{0}^{0}={1\over 8\pi}\varPhi^{;a}\varPhi_{;a}+\varPhi J\,.
\ee
This expression for the energy can be rewritten as
\be\begin{split}
{\mathcal E}&=-{1\over 2}\int dx\,\sqrt{g}\,\alpha\,J\varPhi-{1\over
8\pi}\int
dx\,\partial_{a}\left(\sqrt{g}\,\alpha\,\varPhi\,
\varPhi^{;a}\right)
\, .
\end{split}\ee
The last term is proportional to the surface integral of $\varPhi\,
\varPhi^{;a} $ over the boundary.  It vanishes because of the boundary
conditions at infinity and on the horizon.

Similarly to the electromagnetic case it is useful to introduce another field
variable $\varphi$
\be\label{varphi}
\varPhi=\alpha^{-1/2}\,\varphi\,.
\ee
The field $\varphi$ satisfies the equation
\be\begin{split}\label{eqphi}
(\lap  +V)\,\varphi&=-4\pi j\,,
\end{split}\ee
\be\begin{split}\label{V0}
V&={(\nabla\alpha)^2\over
4\alpha^2}-{\lap\alpha\over 2\alpha}\equiv -{\lap(\alpha^{1/2})\over
\alpha^{1/2}}\,,
\end{split}\ee
\be\begin{split}\label{j}
j&=\alpha^{1/2} J \,.
\end{split}\ee
Here
\be
\lap=g^{ab}\nabla_a\nabla_b\,.
\ee

\subsection{Self-energy}

In terms of this field the energy can be rewritten in the form
\be\begin{split}\label{EEE}
{\mathcal E}&=-{1\over 8\pi}\int
dx\,\sqrt{g}\,g^{ab}\left(\varphi_{,a}-{\alpha_{,a}\over
2\alpha}\varphi\right)\left(\varphi_{,b}-{\alpha_{,b}
\over
2\alpha}\varphi\right)
\, .
\end{split}\ee
This expression for the energy can be interpreted as the Euclidean action of a
$(D-1)$-dimensional
scalar field $\varphi$ interacting with the external dilaton  field $\alpha$.
One can use this  analogy to reformulate the problem of calculation of the
self-energy in terms of the Euclidean quantum field theory defined on
$(D-1)-$dimensional space and in the presence of the external dilaton field.

For a point scalar charge located at $y$ the charge distribution reads
\be\label{J}
J=q{1\over\sqrt{g}}\delta (x-y)\,.
\ee
Thus, the self-energy function of
point scalar charges can be written in the form \cite{Frolov:2012zd}
\be\label{E0}
{\mathcal E}=-2\pi q^2 \alpha(y)\,{G}(y,y)=-2\pi q^2
\alpha(y)\,\langle\varphi^2\rangle\,.
\ee
Here ${G}(y,y)$ is the coincidence limit $x,x'\rightarrow y$ of
the regularized Green function.
The Green function
${G}$ corresponds to the operator  \eq{eqphi}
\be\label{FG}
(\lap  +V)\,{G}(x,x')=-\delta(x,x')\, .
\ee

For point sources energy function ${\mathcal E}$ diverges. To deal
with this divergence one has to use some regularization and renormalization
schemes. Following the same arguments as in electromagnetic case we obtain
\be\begin{split}
{\mathcal E}_{\inds{\epsilon}}&=-2\pi q^2\, \lim_{x,x'\rightarrow y}
\sqrt{\alpha(x)\alpha(x')}\,{G}_{\inds{\epsilon}}(x,x')\,.
\end{split}\ee

\subsection{Self-force}

The self-force of the static scalar charge distribution $J$ in the
static spacetime can be defined
\cite{Quinn:2000wa,Isoyama:2012in} exactly in the same way as that of the
electric charge distribution \eq{defforce}-\eq{forcedensity}.
\be
f_{a}(y)={1\over\alpha(y)}\int_{\Sigma} g_{a}{}^b(y,x)\,
\mathrm{f}_b(x)\,\alpha(x)\sqrt{g(x)} \,dx\,,
\ee
\be
\mathrm{f}_\alpha=-T_{\alpha\beta}{}^{;\beta}\,.
\ee
The stress-energy tensor of the scalar charge \eq{ScalarTmumu} without
mass terms reads
\be\label{Tab_scalar}
T^{\mu\nu}={1\over 4\pi}\left(\varPhi^{;\mu} \varPhi^{;\nu}-{1\over 2}
g^{\ins{D}}{}^{\mu\nu}\varPhi^{;\alpha}\varPhi_{;\alpha}\right)
-u^\mu u^\nu\,\varPhi J\,.
\ee
The divergence of the stress-energy tensor gives the density of the self-force
\be
\mathrm{f}_\alpha=\varPhi
J\,w_\alpha+J(\delta_\alpha^\beta+u_\alpha u^\beta)\nabla_\beta\varPhi\,,
\ee
where
\be
w_\alpha=u^\sigma\nabla_\sigma u_\alpha
\ee
is the acceleration vector corresponding to the worldline with the velocity
$u^\alpha$.

We define the self-force density of the scalar charge distribution as the
right-hand side of the equation of motion \eq{fscalar}
\be\label{scalarequation}
\mu\,w_\alpha = \mathrm{f}_\alpha\,.
\ee
Here $\mu$ is the mass density of the source and
\be\label{fscalardensity}
\mathrm{f}_\alpha=J
(\delta_\alpha^\beta+u_\alpha
u^\beta)\nabla_\beta\varPhi+w_\alpha\,\varPhi J\,.
\ee
This force, evidently, is orthogonal to the velocity $u^\alpha$ of the flux. So,
for the $T^{\mu\nu}$ in the form \eq{Tab_scalar}  the problem of
dependence of the mass of the point scalar charge on its proper
time, discussed in \cite{Quinn:2000wa}, does not appear.

Note that the force density \eq{fscalardensity} is the sum of terms of two
kinds. The first one is determined by a gradient of the scalar field, while
the second term in $\mathrm{f}_\alpha$ is proportional to the
acceleration vector and to the value of the scalar field. The second term has
the structure similar to the contribution of the the mass density, though it
depends on the value of the scalar field at the position of the charge. This
similarity is sometimes used to move this term to the left-hand side of the
motion equation but redefine the inertial mass density of the charge and the
force (see, e.g., \cite{Poisson:2011nh} and references therein) as
\be\label{tildemu}
\tilde{\mu}=\mu-J\varPhi\,,
\ee
\be\label{tildef1}
\tilde{\mathrm{f}}_\alpha=\mathrm{f}_\alpha-w_\alpha\,\varPhi J\,.
\ee
In this representation the force $\tilde{\mathrm{f}}_\alpha$ depends only on
$\nabla_\beta\varPhi$
\cite{Quinn:2000wa,Poisson:2011nh,Beach:2014aba,Zimmerman:2014uja}
\be\label{tildef2}
\tilde{\mathrm{f}}_\alpha=J (\delta_\alpha^\beta+u_\alpha
u^\beta)\nabla_\beta\varPhi\,,
\ee
but in the general case the `inertial' mass $\tilde{\mu}$ is position dependent.

In this paper we use the motion equations
Eqs.(\ref{scalarequation}),(\ref{fscalardensity}),(\ref{fscalar}) in their
original form, that is without any redefinition of the mass and the force. This
choice guarantees an agreement between the change of the self-energy function
of
the
charge and the work of the self-force during adiabatic displacement of the
charge. At the end of this section we provide the results for both kinds of
forces.

In the case of a static charge in the static spacetime \eq{metric} the
expression
for
self-force reads
\be\begin{split}
f_{a}(y)&=-{1\over\alpha(y)}\int_{\Sigma}
g_{a}{}^b(y,x)T_{b\beta}{}^{;\beta}(x)\,\alpha(x)\sqrt{g(x)} dx
\\
&={1\over\alpha(y)}\int_{\Sigma} g_{a}{}^b(y,x) \left(J \nabla_{b}\varPhi +J
u_{b}u^{\sigma}\nabla_{\sigma}\varPhi + w_b\, J \varPhi
\right)\,\alpha(x)\sqrt{g(x)} dx\,.
\end{split}\ee
Its velocity is proportional to the Killing
vector, while the acceleration is $w_\mu=\alpha^{-1}\partial_\mu \alpha$. The
time component of the self-force and $\nabla_0\varPhi$ vanish, and we obtain
\be
f_{a}(y)={1\over\alpha(y)}\int_{\Sigma} dx
\sqrt{g}\alpha g_{a}{}^b(y,x)\left(J \nabla_{b}\varPhi + w_b\, J
\varPhi\right)\,.
\ee
Substituting \eq{varphi} and \eq{j} in this relation one gets
\be\begin{split}
f_{a}(y)&={1\over\alpha(y)}\int_{\Sigma} dx\,
\sqrt{g} g_{a}{}^b(y,x)\left(j \nabla_{b}\varphi + {1\over 2} w_b\, j
\varphi\right)\\
&={4\pi\over\alpha(y)}\int_{\Sigma} dxdx'\,
\sqrt{g(x)}\sqrt{g(x')}\,j(x)j(x')g_{a}{}^b(y,x)
\left({\partial\over\partial x^b}G(x,x')
+{1\over 2}w_b\,G(x,x')\right)\\
&={4\pi\over\alpha(y)}\int_{\Sigma} dxdx'\,
\sqrt{g(x)}\sqrt{g(x')}\,{j(x)j(x')\over \sqrt{\alpha(x)}\sqrt{\alpha(x')}}
g_{a}{}^b(y,x){\partial\over\partial
x^b}\left(\sqrt{\alpha(x)}\sqrt{\alpha(x')}\,G(x,x')\right)\,.
\end{split}\ee
For a point charge located at $y$ the charge density distribution is
\be
j=q{\sqrt{\alpha}\over\sqrt{g}}\delta (x-y)\,.
\ee
So that one finally obtains
\be\begin{split}
f_{a}(y)&={4\pi q^2\over\alpha(y)}
{\partial\over\partial
x^a}\left(\sqrt{\alpha(x)}\sqrt{\alpha(x')}\,G(x,x')\right)\Big|_{x=y,x'=y}\,.
\end{split}\ee

The derivative has to be taken before placing points $x$ and $x'$ on the
worldline of the charge. This divergent expression has to be
regularized. Following the lines of the electromagnetic case
one can rewrite it formally as
\be
f_{\inds{\epsilon}}{}_{a}(y)
=2\pi q^2 {1\over\alpha(y)} {\partial\over\partial y^{a}}
\left(\alpha(y)\,G_{\inds{\epsilon}}(y,y)\right)\,.
\ee

Comparing this formula with the expression for the self-energy function \eq{E0}
one can see that
\be\label{Ef}
f_{a}=-{1\over\alpha}\partial_{a}{\mathcal E}\,.
\ee
Thus, the relation between the self-force and the variation of
the self-energy function over the position of the charge is the same as in the
electromagnetic case. This fact is not surprising but it's a good test of the
validity of derivation of the self-force\footnote{
In the articles
\cite{Quinn:2000wa,Poisson:2011nh,Beach:2014aba,Zimmerman:2014uja}
the modified self-force $\tilde{f}_\mu$  has been used. It can be
derived
by integrating densities Eqs.(\ref{tildef1}) and (\ref{tildef2}). The result of
computations can be presented in the form
\be\label{Etildef}
\tilde{f}_\mu=f_\mu+{2\over \alpha} w_\mu\,
{\mathcal
E}=-{1\over
\alpha}\left(\partial_\mu {\mathcal E} - 2 w_\mu {\mathcal E}\right)\,.
\ee
This formula is applicable to static scalar charges in arbitrary static
spacetimes.
}.


\section{Scalar charge in the Rindler spacetime}\label{ScalarRindler}

For the scalar field in the $D-$dimensional Rindler space we have to find the
regularized Green function and the
heat kernel for the operator $\lap +V$, where $\lap$ is the flat
$(D-1)$-dimensional
flat Laplace operator and the potential
$V={1/(4 z^2)}$
\be
(\lap+V)\,G(x,x')
=\left[\partial_z^2+\partial_{\boldsymbol{x}_{\perp}}^2+{1\over
4z^2}\right]G(x,x')=-\delta(x,x')\, .
\ee
Choosing the same regularization as for Maxwell field we get
\be
K_{\inds{\epsilon}}(s|x,x')=\rho(s,\epsilon)\,K(s|x,x')\hh
\rho(s,\epsilon)=\exp(-{\epsilon^2/4s})\,.
\ee
The regularized static Green function is
\be\label{GKds0}
G_{\inds{\epsilon}}(x,x')=\int_0^{\infty}ds\,K_{\inds{\epsilon}}
(s|x,x')\,.
\ee
One can check that the solution for the regularized static heat
kernel is
\be
K_{\inds{\epsilon}}=\sqrt{zz'}\,{2\pi\over(4\pi s)^{D/2}}\,
e^{\displaystyle{-{z^2+z'{}^2+\boldsymbol{x}_{\perp}^2+\epsilon^2\over
4s}}}\,I_0\left({}zz'\over
2s\right)\,,
\ee
where $I_0$ is the Bessel function and
$
\boldsymbol{x}_{\perp}^2 \equiv g_{ij}(x^i-x'{}^i)(x^j-x'{}^j)
$.

The regularized Green function is the integral \eq{GKds0} of the regularized
heat kernel over the proper time
\be\label{ScalarG}
{G}_{\inds{\epsilon}}={\sqrt{zz'}\over 2\pi^\beta}\,I_{0}^{\beta}
\hh
\beta={D\over 2}-1\,,
\ee
where $I_{\nu}^{\eta}$ is defined in \eq{Prudnikov}.
The regularized static scalar Green function can be written in the exact
closed form
\be\label{Gepsilon}
{G}_{\inds{\epsilon}}={\Gamma\left(\beta\right)
\over 2\pi^{\beta}}
{(zz')^{1/2}\over (R^2+\epsilon^2)^{\beta}}\,{F}
\left(\beta,{1\over 2};1;-{4zz'\over R^2+\epsilon^2}\right)
\,,
\ee
where $R$ is given by \eq{R}.

Note that the static Green function for the original scalar field $\Phi$
(see \eq{BoxPhi}), differs
from ${G}_{\inds{\epsilon}}$ by the factor $(zz')^{-1/2}$
\be\label{BG}
\mathbb{G}(x,x')=\langle
\Phi(x)\Phi(x')\rangle=(zz')^{-1/2}\,{G}(x,x')\,.
\ee
Along the lines of the electromagnetic case, one can obtain the general
solution for the Green function $\mathbb{G}(x,x')$
from \eq{BG} if one adds zero-mode solutions parametrized by three arbitrary
constants
$
C_0+C_1\,\ln(zz')+C_2\,\ln z\ln z'\,.
$
This general solution reads
\be\label{BGG}
\mathbb{G}(x,x')={\Gamma\left({D\over 2}-1\right)
\over 2\pi^{{D\over 2}-1}}
{1\over R^{D-2}}\,{F}
\left({D\over 2}-1,{1\over 2};1;-{4zz'\over
R^2}\right)+C_0+C_1\,\ln(zz')+C_2\,\ln z\ln z'\,.
\ee
The three arbitrary constants $C_0,C_1,C_2$ are to be
fixed by the boundary conditions when $x$ or $x'$ are at the horizon and at
infinity. Analysis, similar to that of the solution for the vector potential of
the electric charge, shows that regular at the horizon and vanishing at
infinity scalar field configuration corresponds to $C_0=C_1=C_2=0$.
One can check that ${G}_{\inds{\epsilon}}(x,x')$ given by \eq{Gepsilon} 
vanishes  when
one of the arguments, e.g., $x$, is placed on the horizon. It means that it
satisfies zero Dirichlet boundary conditions on the Rindler horizon.
The static Green functions for particular dimensions of the spacetime are given
in the Appendix~\ref{GreenFunctions}. Their asymptotics at
$\epsilon\rightarrow
0$ are in Appendix~\ref{ScalarField}.


\section{Scalar self-energy and self-force: Results}\label{ScalarResults}

\subsection{Divergent part of the static Green function}

To renormalize the self-energy one has to subtract the UV divergent parts from
the Green function \eq{ScalarG}. The local structure of divergences is given
by the first terms in expansion \eq{DeWitt} of the heat kernel over the proper
time parameter. In the case of the scalar charge in the Rindler spacetime the
corresponding Schwinger-DeWitt coefficients are
\be
a_0(x,x)=1\hh
a_1(x,x)=V={1\over 4\,z^2}\hh
a_2(x,x)={1\over 2}V^2+{1\over 6}\lap V={9\over 32\,z^4}\,.
\ee

The residual finite value for quantum fluctuations of the field $\varphi$ at the
point $y$ reads
\be
\langle\varphi^2\rangle^{\ins{res}}={G}_{\inds{\epsilon}}(y,y)-{G}_{
\inds{\epsilon}}{}{}^{\ins{div}}(y,y)\,,
\ee
which enters the formula \eq{E0} for the self-energy of the charged particle.
For different spacetime
dimensions (see Appendix~\ref{ScalarField}) we finally obtain the
self-energy function of the scalar charge moving with the constant acceleration.

\subsection{Self-force}

The divergent and finite parts of the residual self-force acting on the scalar 
charge can be computed along the same lines as in the electromagnetic case 
(Section~\ref{MaxwellSelf-force}).
\be\label{ff0}
m_0 a_{\mu}=F_{\mu}^\ins{ext}+f_{\mu} \hh
f_{\mu}=f_{\mu}^\ins{div}+f_{\mu}^\ins{res}\, .
\ee
Here $f_{\mu}^\ins{div}$ is a divergent in the limit $\epsilon\to 0$ part of 
the self-force and $f_{\mu}^\ins{res}$ is the finite residual part of the 
self-force. They can be derived from the corresponding divergent and residual 
finite parts of the self-energy using the relations (see also \eq{fepsilon})
\be
f_{\inds{\epsilon}}{}_{a} =-{1\over \alpha}{\partial\over\partial 
y^{a}}{\mathcal E}_{\inds{\epsilon}}\hh
f_{a}^{\ins{div}}=-{1\over \alpha}{\partial\over\partial 
y^{a}}{\mathcal E}^{\ins{div}}\hh
f_{a}^{\ins{res}}=-{1\over \alpha}{\partial\over\partial 
y^{a}}{\mathcal E}^{\ins{res}}\,.
\ee
Note that in the Rindler spacetime  $\alpha=az$. It is equal to one at the 
point of observation.

It should be emphasized that in this approach the bare mass $m_0$, or its 
renormalized version $m$ (see \eq{m0m}), are constants. In the literature on 
the self-force of scalar charges 
\cite{Quinn:2000wa,Poisson:2011nh,Beach:2014aba,Zimmerman:2014uja,
Gal'tsov:2004qz,Galtsov:2007zz} the same motion equation \eq{ff0} is often 
presented in the form where some specific, proportional to the acceleration,  
part of $f_a$ is moved to the left-hand side of the motion equation and 
included into the redefinition of the {\it inertial} mass (see \eq{tildemu}, 
\eq{tildef1}, \eq{tildef2}), which then becomes dependent on the position of the 
scalar charge. We denote the corresponding remaining part of the self-force on 
the right-hand side of the motion equation as $\tilde{f}_a$. It is related to 
the self-energy function as 
\be
\tilde{f}_{a}^{\ins{div}}=-{1\over \alpha}\left({\partial\over\partial 
y^{a}}{\mathcal E}^{\ins{div}}-2 w_a {\mathcal 
E}^{\ins{div}}\right)\hh
\tilde{f}_{a}^{\ins{res}}=-{1\over \alpha}\left({\partial\over\partial 
y^{a}}{\mathcal E}^{\ins{res}}-2 w_a {\mathcal 
E}^{\ins{res}}\right)\,.
\ee

The divergent part of the self-force has the following structure
\be\label{PP0}
f_{\mu}^\ins{div}=q^2 a_{\mu} P(a)\hh
P(a)=\sum_{k=0}^{k_D} c_D^{(2k)} {a^{2k}\over \epsilon^{D-3-2k}}\, .
\ee
Here $k_D=(D-3)/2$ for odd $D$ and $k_D=D/2-2$ for even $D$.
For odd $D$ in the last term of this expression one should make a substitution
\be
{1\over \epsilon^0}\to \ln(\epsilon\tilde{\lambda})\hh \tilde{\lambda}=\lambda 
e^{\gamma}/2\, ,
\ee
where $\lambda$ is the infrared cut-off. The numerical coefficients 
$c_D^{(2k)}$ are

\bigskip

\hskip 2cm
\begin{tabular}{|r|l|l|l|}
  \hline
  $D=4 \phantom{\displaystyle{1\over1}}$& $c_4^{(0)}={1\over 2}$& & \\
  \hline 
  $D=5\phantom{\displaystyle{1\over1}}$ & $c_5^{(0)}={1\over 2\,\pi}$& 
$c_5^{(2)}={1\over 16\,\pi}$&\\
  \hline 
  $D=6\phantom{\displaystyle{1\over1}}$ & $c_6^{(0)}={1\over 4\,\pi}$& 
$c_6^{(2)}=-{1\over 32\,\pi}$&\\
  \hline
  $D=7\phantom{\displaystyle{1\over1}}$ & $c_7^{(0)}={1\over 2\,\pi^2}$& 
$c_7^{(2)}=-{1\over 32\,\pi^2}$&$c_7^{(4)}={27\over 512\,\pi^2}$\\
  \hline
  $D=8\phantom{\displaystyle{1\over1}}$ & $c_8^{(0)}={3\over 8\,\pi^2}$& 
$c_8^{(2)}=-{1\over 64\,\pi^2}$&$c_8^{(4)}=-{27\over 
1024\,\pi^2}$\\
  \hline
\end{tabular}
\bigskip

The finite residual part of the self-force has the structure 
\be
f_{\mu}^\ins{res}=q^2 a_{\mu} a^{D-3}[A_D +B_D 
\ln(8\tilde{\lambda}/a)]\, .
\ee
For even $D$ the logarithmic terms are absent and one has $B_D=0$.
Our calculations lead to the following the coefficients $A_D$ and $B_D$ 

\bigskip

\hskip 2cm
\begin{tabular}{|r|l|l|l|}
  \hline
  $D=4 \phantom{\displaystyle{1\over1}}$& $A_4=0$& ${B}_4=0$ \\
  \hline 
  $D=5\phantom{\displaystyle{1\over1}}$ & $A_5={5\over 32\,\pi}$& 
${B}_5=-{1\over 16\,\pi}$\\
  \hline 
$D=6\phantom{\displaystyle{1\over1}}$ & $A_6=0$& 
${B}_6=0$\\
  \hline 
$D=7\phantom{\displaystyle{1\over1}}$ & $A_7={243\over 2048\,\pi^2}$& 
${B}_7=-{27\over 512\,\pi^2}$\\
  \hline 
$D=8 \phantom{\displaystyle{1\over1}}$& $A_8=0$& ${B}_8=0$ \\
  \hline 
\end{tabular}
\bigskip

Similarly we obtain  the coefficients for the divergent and finite 
residual parts of $\tilde{f}_a$

\be\label{PPP0}
\tilde{f}_{\mu}^\ins{div}=q^2 a_{\mu} \tilde{P}(a)\hh
\tilde{P}(a)=\sum_{k=0}^{k_D} \tilde{c}_D^{(2k)} {a^{2k}\over 
\epsilon^{D-3-2k}}\, .
\ee

\bigskip

\hskip 2cm
\begin{tabular}{|r|l|l|l|}
  \hline
  $D=4 \phantom{\displaystyle{1\over1}}$& $\tilde{c}_4^{(0)}=-{1\over 2}$& & \\
  \hline 
  $D=5\phantom{\displaystyle{1\over1}}$ & $\tilde{c}_5^{(0)}=-{1\over 2\,\pi}$& 
$\tilde{c}_5^{(2)}={3\over 16\,\pi}$&\\
  \hline 
  $D=6\phantom{\displaystyle{1\over1}}$ & $\tilde{c}_6^{(0)}=-{1\over 4\,\pi}$& 
$\tilde{c}_6^{(2)}=-{3\over 32\,\pi}$&\\
  \hline
  $D=7\phantom{\displaystyle{1\over1}}$ & $\tilde{c}_7^{(0)}=-{1\over 
2\,\pi^2}$& 
$\tilde{c}_7^{(2)}=-{3\over 32\,\pi^2}$&$\tilde{c}_7^{(4)}={45\over 
512\,\pi^2}$\\
  \hline
  $D=8\phantom{\displaystyle{1\over1}}$ & $\tilde{c}_8^{(0)}=-{3\over 
8\,\pi^2}$& 
$\tilde{c}_8^{(2)}=-{3\over 64\,\pi^2}$&$\tilde{c}_8^{(4)}=-{45\over 
1024\,\pi^2}$\\
  \hline
\end{tabular}
\bigskip

The finite residual part of the tilde-self-force has the structure 
\be\label{fmures0}
\tilde{f}_{\mu}^\ins{res}=q^2 a_{\mu} a^{D-3}[\tilde{A}_D +\tilde{B}_D 
\ln(8\tilde{\lambda}/a)]\, .
\ee

\bigskip

\hskip 2cm
\begin{tabular}{|r|l|l|l|}
  \hline
  $D=4 \phantom{\displaystyle{1\over1}}$& $\tilde{A}_4=0$& $\tilde{B}_4=0$ \\
  \hline 
  $D=5\phantom{\displaystyle{1\over1}}$ & $\tilde{A}_5={11\over 32\,\pi}$& 
$\tilde{B}_5=-{3\over 16\,\pi}$\\
  \hline 
$D=6\phantom{\displaystyle{1\over1}}$ & $\tilde{A}_6=0$& 
$\tilde{B}_6=0$\\
  \hline 
$D=7\phantom{\displaystyle{1\over1}}$ & $\tilde{A}_7={381\over 2048\,\pi^2}$& 
$\tilde{B}_7=-{45\over 512\,\pi^2}$\\
  \hline 
$D=8 \phantom{\displaystyle{1\over1}}$& $\tilde{A}_8=0$& $\tilde{B}_8=0$ \\
  \hline 
\end{tabular}
\bigskip

One can see an amazing coincidence of electromagnetic residual self-forces 
${f}_{\mu}^\ins{div}$ and ${f}_{\mu}^\ins{res}$ (see \eq{PP}, \eq{fmures}) with 
the scalar ones $\tilde{f}_{\mu}^\ins{div}$ and $\tilde{f}_{\mu}^\ins{res}$ 
(see \eq{PPP0}, \eq{fmures0}).

\subsection{Self-energy}

In even dimensions $D=4,6,8$, etc. self-energy function vanishes.
In odd-dimensional spacetimes we have
\be
{\mathcal E}=-{q^2\,a\over 16\pi z}
\left[\ln\left({4\lambda z}\right)+\gamma-{3\over 2}\right]\hh D=5\,,
\ee
\be
{\mathcal E}=-{9\,q^2\,a\over 512\pi^2 z^3} \left[\ln\left({
4\lambda z}\right)+\gamma-{23\over 12}\right]\hh D=7\,.
\ee
As earlier the self-energy functions ${\mathcal E}^{\ins{res}}(x_o,y)$ depend
on two
points, a position of the charge $y^a$ and the point of normalization of the
Killing
vector $x_o^a$. We put $x_o^a=(a^{-1},x^i=0)$ and $y^a=(z,y^i=0)$ and wrote the
residual self-energy functions as functions two parameters $a$ and $z$. For the
residual self-energy
one has ${E}^{\ins{res}}(a^{-1})={\mathcal E}^{\ins{res}}(a^{-1},a^{-1})$.
The residual self-force of the charge is calculated at the point $z=a^{-1}$.

\subsection{Summary of results}

The results of the calculations of the finite residual self-energy 
${E}^{\ins{res}}$ 
and self-forces ${f}_a^{\ins{res}}$ and $\tilde{f}_a^{\ins{res}}$ can be
summarized as follows:

\begin{itemize}
\item Four dimensions. The self-energy and self-forces vanish.
\be
{E}^{\ins{res}}=0 \hh
f_z^{\ins{res}}=0 \hh
\tilde{f}_z^{\ins{res}}=0\,.
\ee
\item Five dimensions.
\be
{E}^{\ins{res}}=-{q^2\,a^2\over 16\pi}
\left[\ln\left({4\lambda \over a}\right)+\gamma-{3\over 2}\right]\,,
\ee
\be
f_z^{\ins{res}}=-{q^2 a^3\over 16\pi}
\left[\ln\left({4\lambda \over a}\right)+\gamma-{5\over 2}\right]\,,
\ee
\be
\tilde{f}^{\ins{res}}_z=-{3 q^2 a^3\over 16\pi}
\left[\ln\left({4\lambda \over a}\right)+\gamma-{11\over 6}\right]\,.
\ee
\item Six dimensions. The self-energy and self-force vanish
\be
E^{\ins{res}}=0 \hh
f_z^{\ins{res}}=0\hh
\tilde{f}_z^{\ins{res}}=0\,.
\ee
\item Seven dimensions.
\be
{E}^{\ins{res}}=-{9\,q^2\,a^4\over 512\pi^2} \left[\ln\left({
4\lambda z}\right)+\gamma-{23\over 12}\right]\,,
\ee
\be
f_z^{\ins{res}}=-{27\,q^2 a^5\over 512\pi^2} \left[\ln\left({
4\lambda \over a}\right)+\gamma-{9\over 4}\right]\,,
\ee
\be
\tilde{f}_z^{\ins{res}}=-{45\,q^2 a^5\over 512\pi^2} \left[\ln\left({
4\lambda \over a}\right)+\gamma-{127\over 60}\right]\,.
\ee
\item Eight dimensions
\be
{E}^{\ins{res}}=0\hh
f_z^{\ins{res}}=0\hh
\tilde{f}_z^{\ins{res}}=0\,.
\ee
\end{itemize}

Similar to the electric charge the self-force for the scalar charge vanishes in
even dimensions.In odd dimensions the self-energy and the self-force
depend on the IR cutoff $\lambda$ that needs to be chosen from
physical arguments (see discussion of this problem, e.g., in
\cite{Beach:2014aba}).


\section{Discussion}\label{Discussion}

In the present paper we calculated a self-energy  and a self-force  for a point
charge in a higher-dimensional static homogeneous gravitational field.

One of the results, presented in the present paper, seems to be quite
interesting and intriguing. Namely, we demonstrated that in even dimensions
$4\le D\le 8$ the finite residual part of the self-force for both
electric and scalar charge identically vanishes.
We do not know a deep
reason for this, but one might make the conjecture, that is a common
feature of even dimensional spacetimes with the number of dimensions
greater than four.

In 4 dimensions this result can be
expected for the following reason. The invariant residual self-force acting on 
a charge
$e$
near a Schwarzschild or a Reissner-Nordstr\"{o}m black hole of mass $M$ is
\cite{Smith:1980tv,Zelnikov:1982in,Poisson:2011nh}
\be
f^{\ins{res}}=(f^{\ins{self}\mu}f^{\ins{res}}_{\mu})^{1/2}= {e^2 M\over r^3}\,
.
\ee
In order to pass to the Rindler limit in this formula one needs to take the
limit,
when the gravitational radius $r_g=2M$ infinitely grows while the proper 
distance
to
the horizon is fixed. In this limit $f^{\ins{res}}= 0$ that coincides with our
result. In the higher even dimensions the vanishing of the residual self-force 
in the
Rindler
spacetime implies that the corresponding invariant of this self-force for
the charge near  the
black
hole remains finite on the horizon. This result is valid both for electric
and scalar charges.

The results obtained for a static homogeneous gravitational field can be
directly applied without any changes to the case of a uniformly accelerated 
motion of the charge in the higher dimensional Minkowski spacetime. In four 
dimensions the vanishing of the resudual self-force acting on a uniformly 
accelerated charge is well known property (see e.g. 
\cite{Ginzburg:1985,Rohrlich:2008}). The obtained in this paper results indicate 
that a similar property is valid for accelerated scalar and electric charges in 
any even dimensional  Minkowski spacetime.

A natural question arises about the validity of the calculations and
the results. For example, one can ask the question: At what distance from the 
horizon the
residual self-force
of the charge becomes comparable with the gravitational attraction of the
particle with the mass $m$. Let us neglect
for a
moment logarithmic factors. The finite part of the self-force is of the order
of
${f^{\ins{res}}}\sim e^2 a^{D-2}$. On the other hand, in order to provide the
acceleration of a particle of mass $m$, one should apply the force
$f=ma$. They become comparable
$
{f^{\ins{res}}}\sim f\, ,
$
when
\be\label{aaa}
a_*^{D-3}\sim {m\over e^2}\, .
\ee
We denote by $l_*=a_*^{-1}$ a distance to the horizon corresponding to
this solution. We also denote by  $r_m$ a classical radius of the
charged particle
\be
m={e^2\over r_m^{D-3}}\, ,
\ee
then condition \eq{aaa} implies that
\be
l_*\sim r_m\, .
\ee
In other words the induced finite self-force correction, acting on
the charge, is of the same order as the force, which produces the acceleration
(directed towards the horizon \cite{Beach:2014aba})
when the distance to the horizon is small and comparable with the classical
radius of the
charge. One can
expect, that study of such regimes would require more detailed
knowledge concerning the internal structure of the particle, so that
one cannot trust the obtained  results in this range of parameters.
In a way, this problem is  similar to the famous problem of
self-accelerated radiating charged particle motion in the classical
electrodynamics (see, e.g. \cite{Landau:1982dva,Rohrlich:2008,Gralla:2009md}).

As we have already told the static homogeneous gravitational field  is
an idealization. The corresponding Rindler metric naturally
served as an approximation for the static metric of a compact object
in a spacetime domain of the size small with respect to the size of
the object (a black hole). An interesting question is how the
expressions for the mass-shift due to the self-interaction for a
particle in the homogeneous gravitational field and near the black
hole are connected. In four-dimensional case this problem was analyzed
long time ago (see, e.g., \cite{Smith:1980tv,Zelnikov:1983,Zelnikov:1982in}). 
The self-energy shift for a
charged particle near a black hole can be written in the form
\be\label{bh}
E^\ins{res}={1\over 2}e^2\, a\, .
\ee
This expression is similar to \eq{exact}, but differs from it by the
sign. This difference is a result of two factors: (1) the topology of
the black hole exterior differs from the topology of the Rindler space
in the external with respect to the horizon domain; (2) the integral
expression for the electromagnetic field energy is not uniformly
convergent, so that its limit when $M\to\infty$ differs from a similar
integral, in which this limit is performed `inside' the integral. As a
result of the non-trivial spatial topology of a black hole, the
Maxwell equations for a static electric field in its background allow
the existence of a `zero-mode'. It describes a spherically symmetric
electric field, which does not have sources in the black hole
exterior. By adding such a solution one can make a black hole to be
`weakly charged'.
For the self-mass problem one chooses the Green functions so that such
a mode is suppressed and the black hole remains uncharged. Then the electric
potential on the horizon, which is always constant,
does not vanish. In the homogeneous gravitational field such zero-modes
are absent and a regular solution has zero potential at the horizon.
It is interesting, whether a similar phenomenon occurs in the higher
dimensions.

As we mentioned, in the odd dimensional cases the expressions
for the residual self-energy and self-force contain logarithmic factors.
Similar factors were discovered in \cite{Beach:2014aba} for the self-energy
problem near 5-dimensional black hole.
It is quite interesting question, what is a physical meaning of the
cut-off parameter, that enters these expressions. One of the
options is that it might be related with the scale that controls the
validity of the homogeneous gravitational field approximation.

To obtain finite expressions for the self-energy and self-force we
used special prescription for its regularization. This method is quite natural
and basically it is some covariant version of the Hadamard
regularization. However, an interesting question is whether there is an
agreement in the results for different types of such `covariant'
regularizations. One certainly expects that the results obtained
for a covariant regularization are more robust, than, say, for more naive
regularizations, such as extended charged sphere. In the latter case
an inevitable problem is connected with an ambiguity of the charge
density distribution over the `particle'. In higher dimensions this
problem is more severe than in the 4-dimensional case.

In our approach we study a pure classical problem: The Planck constant
does not enter any of the relations. However, it is quite interesting to
consider a similar problem in the quantum field theory. In the four
dimensional case Ritus \cite{Ritus:1978cj,Ritus:1981jm} calculated the shift of
the
self-mass  for an electron moving in the static homogeneous electric
field. His `quantum' results were in good agreement with the classical
calculations (for details, see \cite{Zelnikov:1982in,Zelnikov:1983}). It is
worthwhile to
repeat similar comparison in higher dimensions.


\acknowledgments{
The authors are grateful to Eric Poisson for inspiring discussions. The
authors also are grateful to the referee for concrete and constructive
suggestions of a better form of presenting the results.
This work was partly supported  by  the Natural Sciences and Engineering
Research Council of Canada. The authors are also grateful to the Killam Trust
for its financial support.}


\appendix


\section{Motion of a continuous charged
medium in a curved spacetime}\label{Electric_medium}

\subsection{Fock's approach}

In the present paper we focus on the properties of point charges. However, it
is
instructive to consider this case as a special limit of continuous media, when
its
density takes the form of the delta-like distribution. Using this approach we
derive
in this appendix the equation of motion and the stress-energy tensor for
charged
media moving in a curved spacetime. For the electrically charged media we
reproduce
well known results. For the case of scalar charge we derive a consistent set of
equations which have quite interesting form and allows us to analyze the motion
of
scalar charges in a curved spacetime in a self-consistent form.
To derive of the equation of motion for distributions of
electric or scalar charges we use n approach developed by Fock
\cite{Fock1964theory}
for the electrically charged media.
We assume the flow lines of the media do not intersect, so that there are no
caustics.
In such a case the flow lines determine a spacetime foliation by 1-dimensional
lines,
which can be parametrized as follows
\be\label{chi}
X^{\alpha}=\chi^\alpha(\lambda,y^b)\, .
\ee
For fixed value of the parameters $y^b$, \eq{chi} defines a single flow line
with
$\lambda$ as a `time' parameter along it. We denote a partial derivative along
the
flow line with respect to $\lambda$ by a dot: $(...)^.=\partial_{\lambda}(...)$.
The velocity of the motion of the element of the media along its flow line is
\be
u^{\alpha}=\dot{\chi}^{\alpha}/L\hh
L=\sqrt{-g_{\mu\nu}\dot{\chi}^\mu \dot{\chi}^\nu}\hh u_{\alpha}u^{\alpha}=-1\, .
\ee
We assume that the proper  mass, $\mu(X)$, as well as of the electric and
scalar
charge, $\mu(X)$ and $J(X)$, within a given pencil of flow lines remain
constant, so
that the following continuity equations are automatically satisfied
\cite{Fock1964theory}.
\be
\nabla_\alpha( \mu u^\alpha)=0\,,
\ee
\be
\nabla_\alpha( J u^\alpha)=0\,,
\ee
\be
\nabla_\alpha( \rho u^\alpha)\equiv \nabla_\alpha( J^\alpha)=0\,.
\ee
In the case of electric charges the conservation of the electric current
$J^\mu$ also follows from the Maxwell equations.

Variation of the flow lines, which is the variation of the function
$\chi^\alpha$, is described by displacement vector
\be
\eta^\alpha=\hat{\delta} X^\alpha=\hat{\delta} \chi^\alpha(\lambda,y^b)
=\partial_{y^b}\chi^\alpha(\lambda,y^b) \delta y^a\,.
\ee
Consider the variation $\tilde{\delta}$ due to the change in the form of the
functions $\chi^\alpha$.
\be
u^\mu (X^\alpha+\eta^\alpha)=u^\mu(X^\alpha)+\tilde{\delta} u^\mu\,,
\ee
\be
\hat{\delta} u^\mu=\tilde{\delta} u^\mu-{\partial u^\mu\over \partial
X^\alpha}\eta^\alpha\,.
\ee
Taking into account that
\be
\tilde{\delta} L= L^{-1} \left[
{1\over 2}
{\partial g^{\ins{D}}_{\mu\nu}\over \partial
X^\sigma}\,\eta^\sigma\, \dot{\chi}^\mu \dot{\chi}^\nu+g_{\mu\nu}
\dot{\chi}^\mu\dot{\eta}^\nu
\right]\,,
\ee
one can write
\be
\tilde{\delta}L
=-L u_\sigma u^\epsilon\nabla_\epsilon \eta^\sigma\,.
\ee
Using these variation rules we obtain
\be
\tilde{\delta} u^\alpha=u^\epsilon{\partial\eta^\alpha\over\partial
X^\epsilon}+u^\alpha u_\sigma u^\epsilon\nabla_\epsilon \eta^\sigma
\ee
and, hence,
\be\begin{split}
\hat{\delta} u^\alpha
&=u^\epsilon\nabla_\epsilon\eta^\alpha
-\eta^\epsilon\nabla_\epsilon u^\alpha
+u^\alpha u_\sigma u^\epsilon\nabla_\epsilon \eta^\sigma\,.
\end{split}\ee
The variations of the mass $\mu(X)$ and scalar charge $J(X)$ densities are
\be
{\tilde{\delta}\mu\over\mu}
+\nabla_\sigma\eta^\sigma=-u_\sigma u^\epsilon\nabla_\epsilon\eta^\sigma\,,
\ee
\be
{\tilde{\delta} J\over J}
+\nabla_\sigma\eta^\sigma=-u_\sigma u^\epsilon\nabla_\epsilon\eta^\sigma\,,
\ee
\be
{\tilde{\delta} \rho\over \rho}
+\nabla_\sigma\eta^\sigma=-u_\sigma u^\epsilon\nabla_\epsilon\eta^\sigma\,.
\ee
Finally the variation of the densities take a form
\be
\hat{\delta}
\mu=-\nabla_{\alpha}(\mu\eta^\alpha)
-\mu u_{\sigma}u^{\epsilon}\nabla_{\epsilon}
\eta^{\sigma}\,,
\ee
\be
\hat{\delta} J=-\nabla_{\alpha}(J\eta^\alpha)
-J u_{\sigma}u^{\epsilon}\nabla_{\epsilon}\eta^{\sigma}\,,
\ee
\be
\hat{\delta}
\rho=-\nabla_{\alpha}(\rho\eta^\alpha)
-\rho u_{\sigma}u^{\epsilon}\nabla_{\epsilon}
\eta^{\sigma}\,.
\ee
Here are the other useful variations
\be\label{delta_mu}
\hat{\delta}(\mu u^\alpha)=\nabla_{\sigma}(\mu u^\sigma\eta^\alpha-\mu
u^\alpha\eta^\sigma )\,,
\ee
\be\label{delta_J}
\hat{\delta}(J u^\alpha)=\nabla_{\sigma}(J u^\sigma\eta^\alpha-J
u^\alpha\eta^\sigma )\,,
\ee
\be\label{delta_rho}
\hat{\delta}(\rho u^\alpha)=\nabla_{\sigma}(\rho u^\sigma\eta^\alpha-\rho
u^\alpha\eta^\sigma )\,.
\ee

\subsection{Electrically charged media}

\subsubsection{Equations of motion}

Let us consider the flow of electrically charged particles with the current
density
$J^\mu(X)=\rho(X)u^\mu$ and  mass density $\mu(X)$.
The action for the Maxwell field including an interaction term with the current
of massive particles reads
\be
I=-{1\over 16\pi}\int dX\,\sqrt{-g^{\ins{D}}}\,F^{\mu\nu}F_{\mu\nu}+\int
dX\,\sqrt{-g^{\ins{D}}}\,A_{\mu}\,\rho\,u^{\mu}
-\int dX\,\sqrt{-g^{\ins{D}}} \,\mu \,.
\ee
Using the variational rule \eq{delta_rho} derived in this appendix,
the continuity equation, and
\be
\hat{\delta}
 \int dX\,\sqrt{-g^{\ins{D}}} \,\mu = \int dX\,\sqrt{-g^{\ins{D}}}
\,\hat{\delta}\mu
-\int dX\,\sqrt{-g^{\ins{D}}}\, \mu\, w_\sigma\eta^\sigma
\ee
we get
\be\begin{split}
\hat{\delta} I&=-{1\over 4\pi}\int
dX\,\sqrt{-g^{\ins{D}}}\,\hat{\delta}A_\mu\left(F^{\mu\sigma}{}_{;\sigma}-4\pi
\rho u^\mu\right)\\
&-\int dX\,\sqrt{-g^{\ins{D}}}\eta^\alpha\left[\mu\,
w_\alpha - \rho u^\beta\,F_{\alpha\beta}\right]
\,.
\end{split}\ee
Here $w^\alpha$ is the acceleration vector
\be
w^\alpha=u^\epsilon\nabla_\epsilon u^\alpha\,.
\ee

Thus the equations of motion are
\be
F^{\mu\sigma}{}_{;\sigma}=4\pi J^\mu  \hh J^\mu=\rho u^\mu\, .
\ee
\be\label{felectric}
\mu\,w_\alpha = \mathrm{f}_\alpha\hh
\mathrm{f}_\alpha=F_{\alpha\beta}J^{\beta}\,.
\ee
The vector $\mathrm{f}_\alpha$ defines the local force density acting on the
element
of an electrically charged medium.


\subsubsection{Stress-energy tensor}

We denote the variations over the metric using the symbol $\delta$
\be
{\delta (\mu\sqrt{-g})\over \mu\sqrt{-g}}={(\partial \chi^\mu/\partial
p)(\partial
\chi^\nu/\partial p) \over 2 g_{\alpha\beta}(\partial
\chi^\alpha/\partial p)(\partial \chi^\beta/\partial p)}\delta
g_{\mu\nu}=-{1\over
2}u^\mu u^\nu \delta g_{\mu\nu}\,.
\ee
The stress-energy tensor
$
T^{\mu\nu}={2\over \sqrt{-g}}{\delta I\over\delta g_{\mu\nu}}
$
reads
\be\label{MaxwellTmunu}
T^{\mu\nu}={1\over 4\pi}\left( F^{\mu\sigma}F^{\nu}{}_{\sigma}-{1\over
4}g^{\mu\nu}F^{\alpha\beta}F_{\alpha\beta}\right)+u^\mu u^\nu \,\mu
\,.
\ee
Note that metric variations of the interaction term vanishes.
The divergence of the stress-energy tensor
\be
T_{\alpha\beta}{}^{;\beta}=\mu w_\alpha-F_{\alpha\beta}J^{\beta}
\ee
must vanish for a closed system. It does vanish on the equations of motion
\eq{felectric}.


\subsection{Motion of the media with the scalar charge}

\subsubsection{Equations of motion}

Let us consider the flow of scalar particles with charge distribution
$J(X)$ and  mass density $\mu(X)$ in a curved spacetime.
A minimally coupled massless scalar field $\varPhi(X)$ with is described by an
action
\be
I=-{1\over 8\pi}\int
dX\,\sqrt{-g^{\ins{D}}}\,\varPhi^{;\mu}\varPhi_{;\mu}
+\int dX\,\sqrt{-g^{\ins{D}}}\, J \varPhi
-\int dX\,\sqrt{-g^{\ins{D}}} \,\mu \,.
\ee
Using the variational rules derived in this appendix and
the continuity equations we get
\be
\hat{\delta}
 \int dX\,\sqrt{-g^{\ins{D}}} \,\mu = \int dX\,\sqrt{-g^{\ins{D}}}
\,\hat{\delta}\mu
=\int dX\,\sqrt{-g^{\ins{D}}}\, \mu\, w_\sigma\eta^\sigma
\,,
\ee
\be\begin{split}
\hat{\delta}
 \int dX\,\sqrt{-g^{\ins{D}}} \,J \varPhi &= \int dX\,\sqrt{-g^{\ins{D}}}
\,\varPhi\hat{\delta} J\\
&=\int
dX\,\sqrt{-g^{\ins{D}}}\,J\eta^\sigma\,\left(\varPhi\,
w_\sigma+\nabla_\sigma\varPhi+u_\sigma u^\epsilon\nabla_\epsilon\varPhi \right)
\end{split}
\,,
\ee
where $w^\alpha$ is the acceleration vector
\be
w^\alpha=u^\epsilon\nabla_\epsilon u^\alpha
\,.
\ee

The variation of the action then reads
\be\begin{split}
\hat{\delta} I&={1\over 4\pi}\int
dX\,\sqrt{-g^{\ins{D}}}\,\hat{\delta}\varPhi\left(\varPhi^{;\mu}_{;\mu}+4\pi
J\right)\\
&-\int dX\,\sqrt{-g^{\ins{D}}}\eta^\sigma\left[\left(\mu-\varPhi J\right)\,
w_\sigma -J(\nabla_\sigma\varPhi+u_\sigma u^\epsilon\nabla_\epsilon\varPhi
)\right]
\,.
\end{split}\ee
Thus the equation of motion are
\be
\Box\,\varPhi=-4\pi J\, .
\ee
\be\label{fscalar}
\mu\,w_\alpha = \mathrm{f}_\alpha\hh \mathrm{f}_\alpha=J
(\delta_\alpha^\beta+u_\alpha
u^\beta)\nabla_\beta\varPhi+w_\alpha\,\varPhi J
\,.
\ee
The vector $\mathrm{f}_\alpha$ defines the local force density acting on the
element
of a charged medium.

\subsubsection{Stress-energy tensor}

We denote the variations over the metric using the symbol $\delta$.
\be
{\delta (\mu\sqrt{-g})\over \mu\sqrt{-g}}={(\partial \chi^\mu/\partial
p)(\partial
\chi^\nu/\partial p) \over 2 g_{\alpha\beta}(\partial
\chi^\alpha/\partial p)(\partial \chi^\beta/\partial p)}\delta
g_{\mu\nu}=-{1\over
2}u^\mu u^\nu \delta g_{\mu\nu}
\,.
\ee
The stress-energy tensor
$
T^{\mu\nu}={2\over \sqrt{-g}}{\delta I\over\delta g_{\mu\nu}}
$ is given by the expression
\be\label{ScalarTmumu}
T^{\mu\nu}={1\over 4\pi}\left(\varPhi^{;\mu} \varPhi^{;\nu}-{1\over 2}
g^{\mu\nu}\varPhi^{;\alpha}\varPhi_{;\alpha}\right)+u^\mu u^\nu (\mu-\varPhi J)
\,.
\ee
Note that, in contrast to the Maxwell field, it contains the term $u^\mu
u^\nu\varPhi J$ explicitly depending on the charge density $J$.
The divergence of the stress-energy tensor
\be
T_{\alpha\beta}{}^{;\beta}=(\mu-\varPhi J)w_\alpha-J
(\delta_\alpha^\beta+u_\alpha
u^\beta)\nabla_\beta\varPhi
\ee
must vanish for a closed system. Thus, we re-derive by a different method the
motion equations \eq{fscalar} for the medium.

\section{Green Functions}\label{GreenFunctions}

The regularized static Green functions, both for scalar and electromagnetic
fields, are given by
\be
{G}_{\epsilon}={\sqrt{zz'}\over
2\pi^\beta}\,I_{\nu}^{\beta}\hh
\beta={D\over 2}-1
\,.
\ee
The parameter $\nu=0$ for the scalar field and $\nu=1$ for the Maxwell
field.

\be\begin{split}\label{Prudnikov1}
I^\eta_\nu&=\int_0^{\infty}u^{\eta-1}\,e^{-pu}I_\nu(cu)\,du \\
&=p^{-\eta-\nu}\left({c\over
2}\right)^{\nu}{\Gamma(\nu+\eta)\over\Gamma(\nu+1)}\,{F}\left({
\eta+\nu\over 2},{\eta+\nu+1\over 2};\nu+1;{c^2\over p^2}\right)\\
&=e^{-\pi\nu i/2}\Gamma(\eta+\nu)(p^2-c^2)^{-\eta/2}\,P^{-\nu}_{\eta-1}
\left( { p\over\sqrt { p^2-c^2 } } \right)\,,
\end{split}\ee
where
\be
u={1\over 4s}\hh p=z^2+z'{}^2+\boldsymbol{x}_{\perp}^2+\epsilon^2\hh c=2zz'\,,
\ee
and
\be
\Re (\eta)>0\hh \Re (p)>|\Re (c)|\,.
\ee

For particular dimensions the integrals $I^\eta_\nu$ can be written in terms
of elliptic functions. Let
\be
k=\sqrt{2c\over p+c}\,,
\ee
then
\be
I^{1/2}_0=k\sqrt{2\over \pi c}\, {\mathbf K}(k) \,,
\ee
\be
I^{1}_0={1\over (p^2-c^2)^{1/2}}\,,
\ee
\be
I^{3/2}_0={k^3\over 2(1-k^2)\sqrt{2 \pi c^3}}\, {\mathbf E}(k) \,,
\ee
\be
I^{2}_0={p\over (p^2-c^2)^{3/2}} \,,
\ee
\be
I^{5/2}_0={k^5\over 8(1-k^2)^2\sqrt{2 \pi c^5}}\,
[  2(2-k^2) {\mathbf E}(k)-(1-k^2){\mathbf K}(k)  ] \,,
\ee
\be
I^{3}_0={2p^2+c^2\over (p^2-c^2)^{5/2}} \,,
\ee
\be
I^{1/2}_1={1\over k}\sqrt{2\over \pi c}\, \left[(2-k^2){\mathbf K}(k)-2{\mathbf
E}(k)\right] \,,
\ee
\be
I^{1}_1={c\over (p^2-c^2)^{1/2}(p+(p^2-c^2)^{1/2})} \,,
\ee
\be
I^{3/2}_1={k\over 2(1-k^2)\sqrt{2 \pi c^3}}\, \left[(2-k^2){\mathbf
E}(k)-2(1-k^2){\mathbf
K}(k)\right] \,,
\ee
\be
I^{2}_1={c\over (p^2-c^2)^{3/2}} \,,
\ee
\be
I^{5/2}_1={k^3\over 8(1-k^2)^2\sqrt{2 \pi c^5}}\,
[  2(1-k^2+k^4) {\mathbf E}(k)-(1-k^2)(2-k^2){\mathbf K}(k)  ] \,,
\ee
\be
I^{3}_1={3 p c\over (p^2-c^2)^{5/2}}\,.
\ee


\section{Divergences and finite parts of the
Green functions}\label{Divergencies}

\subsection{Maxwell field}\label{MaxwellField}

In three dimensions we get

\be {G}_{\inds{\epsilon}}^{(3)}=-{1\over 2\pi
}\left[\ln\left({\epsilon\over
8 z}\right)+2\right]+O(\epsilon^2) \,,
\ee
\be
{G}^{(3)}{}^{\ins{div}}=-{1\over 2\pi}[\ln(\epsilon\lambda)-\ln 2+\gamma] \,,
\ee
\be
{G}^{(3)}{}^\ins{ren}={1\over 2\pi}[\ln(\lambda z)+2 \ln 2 +\gamma -2]\,.
\ee

In four dimensions

\be
{G}_{\inds{\epsilon}}^{(4)}={1\over 4\pi \epsilon}
-{1\over 4\pi z} + O(\epsilon) \,,
\ee
\be
{G}^{(4)}{}^{\ins{div}}={1\over 4\pi}\,{1\over\epsilon } \,,
\ee
\be
{G}^{(4)}{}^\ins{ren}=-{1\over 4\pi z}\,.
\ee

In five dimensions

\be
{G}_{\inds{\epsilon}}^{(5)}={1\over 4\pi^2}\left[{1\over \epsilon^2}
+{3\over 8 z^2 } \left(\ln\left({\epsilon\over
8 z}\right) +{5\over 6} \right)\right]
+O(\epsilon^2) \,,
\ee
\be
{G}^{(5)}{}^{\ins{div}}={1\over 4\pi^2} \left[{1\over\epsilon^2}+{3\over 8 z^2}
\left(\ln (\epsilon\lambda)-\ln 2 +\gamma\right) \right] \,,
\ee
\be
{G}^{(5)}{}^\ins{ren}=-{3\over 32\pi^2 z^2}[\ln(\lambda z)+2\ln 2+\gamma-5/6]\,.
\ee

In six dimensions

\be
{G}_{\inds{\epsilon}}^{(6)}={1\over 8\pi^2 \epsilon^3}-{3\over 64\pi^2
z^2\epsilon}+O(\epsilon) \,,
\ee
\be
{G}^{(6)}{}^{\ins{div}}={1\over 8\pi^2 \epsilon^3}-{3\over 64\pi^2
z^2\epsilon} \,,
\ee
\be
{G}^{(6)}{}^\ins{ren}=0\,.
\ee

In seven dimensions

\be
{G}_{\inds{\epsilon}}^{(7)}={1\over 4\pi^3}\left[{1\over
\epsilon^4}-{3\over 16 z^2\epsilon^2}
+{15\over 256 z^4 } \left(\ln\left({\epsilon\over
8 z}\right) +{107\over 60} \right)\right]
+O(\epsilon^2) \,,
\ee
\be
{G}^{(7)}{}^{\ins{div}}={1\over 4\pi^3} \left[{1\over\epsilon^4}-{3\over 16
z^2\epsilon^2}+{15\over 256 z^4}
\left(\ln (\epsilon\lambda)-\ln 2 +\gamma\right) \right] \,,
\ee
\be
{G}^{(7)}{}^\ins{ren}=-{15\over 1024\pi^3 z^4}[\ln(\lambda z)+2 \ln 2
+\gamma-107/60]\,.
\ee

In eight dimensions

\be
{G}_{\inds{\epsilon}}^{(8)}=
{3\over 16\pi^3}\left[{1\over\epsilon^5}
-{1\over 8 z^2\epsilon^3}-{5\over 128 z^4\epsilon}\right]+O(\epsilon) \,,
\ee
\be
{G}^{(8)}{}^{\ins{div}}=
{3\over 16\pi^3}\left[{1\over\epsilon^5}
-{1\over 8 z^2\epsilon^3}-{5\over 128 z^4\epsilon}\right] \,,
\ee
\be
{G}^{(8)}{}^\ins{ren}=0\,.
\ee


\subsection{Scalar field}\label{ScalarField}

In three dimensions we get

\be
{G}_{\inds{\epsilon}}^{(3)}=-{1\over 2\pi } \ln\left({\epsilon\over
8 z}\right)+O(\epsilon^2) \,,
\ee
\be
{G}^{(3)}{}^{\ins{div}}=-{1\over 2\pi}[\ln(\epsilon\lambda)-\ln 2+\gamma] \,,
\ee
\be
{G}^{(3)}{}^\ins{ren}={1\over 2\pi}[\ln(\lambda z)+2 \ln 2 +\gamma]\,.
\ee

In four dimensions

\be
{G}_{\inds{\epsilon}}^{(4)}={1\over 4\pi \epsilon}+O(\epsilon) \,,
\ee
\be
{G}^{(4)}{}^{\ins{div}}={1\over 4\pi\epsilon } \,,
\ee
\be
{G}^{(4)}{}^\ins{ren}=0\,.
\ee

In five dimensions

\be
{G}_{\inds{\epsilon}}^{(5)}={1\over 4\pi^2}
\left[{1\over\epsilon^2}
-{1\over 8 z^2 } \left(\ln\left({\epsilon\over
8 z}\right) +{3\over 2}\right) \right]
+O(\epsilon^2) \,,
\ee
\be
{G}^{(5)}{}^{\ins{div}}={1\over 4\pi^2} \left[{1\over\epsilon^2}-{1\over 8 z^2}
\left(\ln (\epsilon\lambda)-\ln 2 +\gamma\right) \right] \,,
\ee
\be
{G}^{(5)}{}^\ins{ren}={1\over 32\pi^2 z^2}[\ln(\lambda z)+2 \ln 2 +\gamma-3/2]\,.
\ee

In six dimensions

\be
{G}_{\inds{\epsilon}}^{(6)}={1\over 8\pi^2}
\left[{1\over \epsilon^3}
+{1\over 8 z^2\epsilon}\right]+O(\epsilon) \,,
\ee
\be
{G}^{(6)}{}^{\ins{div}}={1\over 8\pi^2}\left[{1\over\epsilon^3 }
+{1\over 8z^2\epsilon} \right] \,,
\ee
\be
{G}^{(6)}{}^\ins{ren}=0\,.
\ee

In seven dimensions

\be
{G}_{\inds{\epsilon}}^{(7)}={1\over 4\pi^3}\left[{1\over
\epsilon^4}+{1\over 16 z^2 \epsilon^2}
-{9\over 256 z^4 } \left(\ln\left({\epsilon\over
8 z}\right) +{23\over 12} \right)\right]
+O(\epsilon^2) \,,
\ee
\be
{G}^{(7)}{}^{\ins{div}}={1\over 4\pi^3} \left[{1\over\epsilon^4}+{1\over 16
z^2\epsilon^2}-{9\over 256 z^4}
\left(\ln (\epsilon\lambda)-\ln 2 +\gamma\right) \right] \,,
\ee
\be
{G}^{(7)}{}^\ins{ren}={9\over 1024\pi^3 z^4}[\ln(\lambda z)+2 \ln 2
+\gamma-23/12]\,.
\ee

In eight dimensions

\be
{G}_{\inds{\epsilon}}^{(8)}={3\over 16\pi^3 \epsilon^5}+{1\over
128\pi^3
z^2\epsilon^3}+ {9\over 2048\pi^3
z^4\epsilon}+O(\epsilon) \,,
\ee
\be
{G}^{(8)}{}^{\ins{div}}={3\over 16\pi^3 \epsilon^5}+{1\over 128\pi^3
z^2 \epsilon^3 }+{9\over 2048\pi^3 z^4\epsilon } \,,
\ee
\be
{G}^{(8)}{}^\ins{ren}=0\,.
\ee


\providecommand{\href}[2]{#2}\begingroup\raggedright\endgroup


\end{document}